\documentclass[12pt]{article}
\pdfoutput=1

\usepackage{latexsym, amsmath, amsfonts, amssymb, braket}
\usepackage{tikz-cd}
\usepackage[framemethod=TikZ]{mdframed}
\usetikzlibrary{decorations.pathmorphing, cd, decorations.markings, calc, patterns, shapes.symbols}
\usetikzlibrary{decorations.pathreplacing, calligraphy}
\usetikzlibrary{arrows}
\usetikzlibrary{shapes}
\usetikzlibrary{matrix}
\usetikzlibrary{positioning}
\usetikzlibrary{shapes.multipart}
\usepackage{diagbox}
\usetikzlibrary{positioning}
\usetikzlibrary{calc}
\usepackage{xstring}
\usepackage{mathtools}

\usepackage{mathrsfs}
\usepackage[american]{babel}
\usepackage{graphicx}
\usepackage{bbm}
\usepackage[nosort]{cite}

\numberwithin{equation}{section}

\usepackage[colorlinks=true, citecolor=blue!90!black, linkcolor=blue!90!black, linktocpage=true, urlcolor=red!70!black]{hyperref}

\usepackage[left=2.5cm,right=2.5cm,top=2.5cm,bottom=3cm]{geometry}
\def\tilde{\widetilde}
\def\t{\tilde}
\def\hat{\widehat}

\def\bar{\overline}
\def\b{\bar}

\def\half{{1 \over 2}}

\newcommand{\C}{{\mathbb C}}
\newcommand{\R}{{\mathbb R}}
\def\P{\hbox{$\mathbb P$}}

\def\CC{{\mathcal C}}

\def\CM{{\mathcal M}}
\def\CN{{\mathcal N}}
\def\CO{{\mathcal O}}

\def\CT{{\mathcal T}}
\def\CU{{\mathcal U}}

\DeclareFontShape{OT1}{cmr}{mx}{n}%
    {<->cmr10}{}
\newcommand{\mytitlefont}{\fontseries{mx}\selectfont}
\DeclareMathAlphabet{\titlemath}{OT1}{cmr}{mx}{n}

\newcommand{\Z}{\mathbb{Z}}

\numberwithin{equation}{section}

\newcommand{\be}{\begin{equation}} \newcommand{\ee}{\end{equation}}
\newcommand{\bea}{\begin{equation} \begin{aligned}} \newcommand{\eea}{\end{aligned} \end{equation}}

\newcommand{\cA}{\mathcal{A}}

\newcommand{\cC}{\mathcal{C}}

\newcommand{\cF}{\mathcal{F}}

\newcommand{\cM}{\mathcal{M}}

\newcommand{\cO}{\mathcal{O}}

\newcommand{\cT}{\mathcal{T}}
\newcommand{\cU}{\mathcal{U}}

\newcommand{\bC}{\mathbb{C}}

\newcommand{\bI}{\mathbb{I}}

\newcommand{\bP}{\mathbb{P}}

\newcommand{\bZ}{\mathbb{Z}}

\def\repa{\raise4pt\hbox{$\square$}\mkern-14mu\raise-4pt\hbox{$\square$}}
\def\repab{\overline{\raise4pt\hbox{$\square$}\mkern-14mu\raise-4pt\hbox{$\square$}\mkern-1mu}}

\DeclareMathOperator{\tr}{tr}

\DeclareMathOperator{\sign}{sign}

\begin{document}

\begin{titlepage}

\begin{center}

~\\[5pt]

{\fontsize{30pt}{0pt} \mytitlefont Symmetry Breaking from \\[5pt] Monopole Condensation in QED$_3$} 

\vskip25pt

Thomas T.~Dumitrescu,  Pierluigi Niro,  and Ryan Thorngren

\vskip20pt

{\it Mani L.\,Bhaumik Institute for Theoretical Physics,\\[-2pt] Department of Physics and Astronomy,}\\[-2pt]
       {\it University of California, Los Angeles, CA 90095, USA}\\[4pt]

\bigskip
\bigskip

\end{center}

\noindent QED in three dimensions with an $SU(2)_f$ doublet $\psi^i$ of massless, charge-1 Dirac fermions (and no Chern-Simons term) has a $U(2) = (SU(2)_f \times U(1)_m)/\mathbb{Z}_2$ symmetry that acts on gauge-invariant local operators, including monopole operators charged under $U(1)_m$. We argue that there are only two plausible IR scenarios: either the theory flows to a CFT with $U(2)$ symmetry (a scenario strongly constrained by conformal bootstrap bounds); or it spontaneously breaks $U(2) \to U(1)$ via the condensation of a monopole operator of smallest $U(1)_m$ charge, which is a $U(2)$ doublet. This leads to three Nambu-Goldstone bosons described by a sigma model into a squashed three-sphere $S^3$ with $U(2)$ isometry. The conventional $SU(2)_f$-triplet order parameter $i \bar \psi \vec \sigma \, \psi$ also gets a vev, exactly aligned with the monopole vev, such that the triplet parametrizes the $\mathbb{CP}^1$ base of the $S^3$ Hopf bundle, with the monopoles providing the $S^1$ fibers.  We recall why this scenario is compatible with the Vafa-Witten theorem. These results are obtained by considering the phase diagram as a function of the fermion triplet mass $\vec m$. We argue that for all $\vec m \neq 0$ there is a Coulomb phase with a weakly-coupled photon, which arises from a suitable monopole vev; taking $\vec m \to 0$ leads to the symmetry-breaking scenario above. Throughout, we give a detailed account of anomaly matching, which leads to a $\theta=\pi$ term in the $S^3$ sigma model. In one presentation, it can be understood as a Hopf term in a suitably gauged version of the $\mathbb{CP}^1$ sigma model.

\vfill

\end{titlepage}

\pagenumbering{arabic}
\setcounter{page}{1}
\setcounter{footnote}{0}
\renewcommand{\thefootnote}{\arabic{footnote}}

{\renewcommand{\baselinestretch}{.88} \parskip=0pt
	\setcounter{tocdepth}{3}

\tableofcontents}


\section{Introduction and Main Results}
\label{section: Introduction}

\subsection{QED$_3$ with~$N_f$ Flavors}
\label{secetion:qed3intro}
Quantum electrodynamics in three spacetime dimensions (QED$_3$) is the theory of a~$U(1)$ gauge field~$a$ (more precisely, $a$ is a Spin$_c$ connection) coupled to~$N_f$ flavors of two-component Dirac fermions~$\psi^i~(i = 1, \ldots, N_f)$ of electric charge~$+1$. This theory arises in many physical contexts; additionally, it has long served as a simpler foil for the dynamics of QCD in three and four spacetime dimensions. We will study QED$_3$ without a Chern-Simons term for~$a$, which (by virtue of the so-called parity anomaly~\cite{Redlich:1983dv, Redlich:1983kn, Niemi:1983rq, Alvarez-Gaume:1983ihn}) is only possible when~$N_f$ is even. The Lagrangian is thus\footnote{~We mostly work in Lorentzian signature with metric~$\eta_{\mu\nu} = (-,+,+)$ and~$\varepsilon^{012} = 1$. The path integral weight is~$\exp(i S)$ with real~$S = \int \mathscr{L}$. With a slight abuse of notation, we interchangeably write terms in~$\mathscr{L}$ as differential forms or scalar densities, even though the latter do not include the volume element. The 3d gamma matrices satisfy~$\lbrace \gamma^\mu,\gamma^\nu \rbrace = 2\eta^{\mu\nu}$ and we choose~$\gamma^\mu=( i\sigma_y, \sigma_z, -\sigma_x )$. We define the Dirac bar as $\bar\psi = \psi^\dagger \gamma^0$, so that $i\Bar\psi \psi$ and $a_\mu\Bar\psi\gamma^\mu\psi$ are Hermitian operators. Later, especially in discussions of anomaly matching or the Vafa-Witten theorem, we will on occasion switch to Euclidean signature. We use summation conventions for all indices, including for~$SU(N_f)$ (anti-) fundamental flavor indices~$i, j = 1, \ldots, N_f$, which are (down) up, respectively; adjoint indices are denoted as~$I, J = 1, \ldots, N_f^2 -1$.}
\begin{equation}\label{introQEDlag}
\mathscr{L} = - {1 \over 2 e^2} f \wedge \star f - i \b \psi_i \gamma^\mu \left(\partial_\mu - i a_\mu\right) \psi^i~, \qquad f = da~. 
\end{equation}
This theory is weakly coupled in the large-$N_f$ limit, where it can be shown to flow to an interacting CFT (without any symmetry breaking)~\cite{Appelquist:1988sr}; as~$N_f$ is lowered, it becomes more strongly coupled. 

A basic question is for what even values of~$N_f$ (if any) the theory no longer flows to a CFT. The problem has been studied with many different methods, including analytical ones (see e.g.~\cite{Pisarski:1984dj, Nash:1989xx, Maris:1996zg, Kubota:2001kk, Fischer:2004nq, DiPietro:2015taa, Giombi:2015haa, Giombi:2016fct, Herbut:2016ide, Kotikov:2016wrb, Gusynin:2016som, DiPietro:2017kcd,  Benvenuti:2018cwd}) and lattice simulations (see e.g.~\cite{Hands:2002dv, Hands:2002qt, Hands:2004bh, Strouthos:2007stc, Karthik:2015sgq, Karthik:2016ppr}). 
A relatively recent development has been the study of these theories using the conformal bootstrap, starting with~\cite{Chester:2016wrc} (see also the reviews~\cite{Poland:2018epd,Rychkov:2023wsd}). Subsequent bootstrap studies of~QED$_3$~\cite{Li:2018lyb, He:2021sto, Albayrak:2021xtd, Li:2021emd} have been accumulating evidence that the theories with~$N_f \geq 4$ seem consistent with an RG flow to a symmetry-preserving CFT; by contrast, this no longer appears likely for the minimal~$N_f = 2$ theory. The scenario of a symmetry-preserving gapless CFT has the appealing feature that it suggests an enhancement of the global symmetry stemming from a conjectured self-duality of the~$N_f = 2$ theory~\cite{Xu:2015lxa} (see also~\cite{Hsin:2016blu,Wang:2017txt,Cordova:2017kue}). Such symmetry enhancement is not expected in any of the symmetry-breaking phases discussed in this paper.

Taking these results seriously, we will assume that the theory with~$N_f = 2$ flavors does not flow to a symmetry-preserving CFT.\footnote{~Additional evidence for this assumption, as well as for the symmetry-breaking scenario described below, was recently discussed in~\cite{Dumitrescu:2025vfp,Dumitrescu:2026vre}.} A logical possibility not strictly ruled out by bootstrap considerations alone is that the IR theory is a fully symmetric, gapped phase (possibly with a TQFT), but this scenario is not compatible with anomaly matching, nor with the other constraints that we establish below. We are therefore inescapably led to consider scenarios with (at least some) spontaneous symmetry breaking.

\subsection{Symmetries and Local Operators in QED$_3$ with~$N_f =2$}

The global symmetries of massless QED$_3$ with~$N_f = 2$ flavors were analyzed in~\cite{Benini:2017dus, Cordova:2017kue} (see section~\ref{section: UV QED} for more details). There is a continuous zero-form symmetry that acts faithfully on gauge-invariant local operators,
\begin{equation}\label{u2symintro}
    U(2) = {SU(2)_f \times U(1)_m \over \Z_2}~.
\end{equation}
We will refer to~$SU(2)_f$ as the flavor symmetry, and to~$U(1)_m$ as the monopole number (or magnetic) symmetry. In addition, there are discrete symmetries: charge-conjugation~$\CC$, and time-reversal~$\CT$. The fermions~$\psi^i~(i = 1, 2)$ in~\eqref{introQEDlag} are~$SU(2)_f$ doublets, but they are not gauge invariant; the gauge-invariant local operators are all bosonic\footnote{~In particular, the theory can be studied on arbitrary (oriented) three-manifolds~$\CM_3$ without choosing a spin structure.} and come in two varieties:
\begin{itemize}
\item{\bf Non-Monopole Operators:} These are not charged under~$U(1)_m$; they are standard gauge-invariant polynomials in the fields and covariant derivatives. An example we will encounter frequently is the fermion bilinear\footnote{~Here~$\vec \sigma$ are the three Pauli matrices~$\sigma^{I = 1,2, 3}$.}
\begin{equation}\label{TripletO}
    \vec \CO = i \b \psi \, \vec \sigma \, \psi~, \qquad (\vec \CO)^\dagger = \vec \CO~,
\end{equation}
which transforms in the triplet representation of~$SU(2)_f$. Due to the quotient in~\eqref{u2symintro}, all~$U(1)_m$-neutral operators furnish genuine~$SO(3)_f = SU(2)_f/\Z_2$ representations.\footnote{~This is due to the fact that the central~$\Z_2 \in SU(2)_f$ acts on the fermions~$\psi^i$ as a gauge transformation.}
\item{\bf Monopole Operators:} These are gauge-invariant local operators that carry non-zero charge~$q_m \in \Z$ under the~$U(1)_m$ symmetry. They are disorder operators, obtained by constraining the dynamical gauge field~$a$ to have a Dirac monopole singularity of charge~$q_m$ at a fixed (Euclidean) spacetime point.\footnote{~Equivalently, they can be defined via radial quantization on~$S^2 \times \R$, with~$q_m$ units of~$a$-flux on~$S^2$ (see for instance~\cite{Borokhov:2002ib}).} In the presence of the fermions~$\psi^i$, the monopoles can acquire~$SU(2)_f$ quantum numbers because they are dressed with fermion zero modes (see section~\ref{section: UV QED} for more details). In particular, the minimal~$q_m = 1$ monopole is a Lorentz scalar that transforms in the~$SU(2)_f$ doublet representation,
\begin{equation}\label{fundmono}
    \CM^i \quad (i = 1, 2)~, \qquad q_m(\CM^i) = 1~. 
\end{equation}
It is therefore in a faithful representation of the~$U(2)$ symmetry in~\eqref{u2symintro}.\footnote{~More generally, monopoles with odd~$q_m$ transform faithfully under~$SU(2)_f$, while monopoles with even~$q_m$ transform faithfully under~$SO(3)_f = SU(2)_f/\Z_2$.} Its Hermitian conjugate will be denoted by~$\bar \CM_i \equiv (\CM^i)^\dagger$. 

\end{itemize}
An important cautionary remark is that we are studying QED$_3$ with compact~$U(1)$ gauge group, i.e.~local monopole operators exist and are acted on by the~$U(1)_m$ symmetry,\footnote{~This should be distinguished from Abelian gauge theory with non-compact gauge group~$\R$, where the monopoles are no longer genuine local operators (though they do exist as local operators attached to topological lines and should therefore not be ignored), and there is no~$U(1)_m$ zero-form symmetry. It should be possible to obtain this theory from the theory with gauge group~$U(1)$ that we are studying by path integrating over flat~$U(1)_m$ connections. This does not change the local dynamics of the theory, though it can have global effects and modify the symmetries.} but we are not adding them to the Lagrangian, which would explicitly break~$U(1)_m$ (as in Polyakov's confinement mechanism~\cite{Polyakov:1976fu}). Given that~$U(1)_m$ is a good symmetry, we can then ask whether or not it is spontaneously broken by a monopole operator (with~$q_m \neq 0$) that acquires a vacuum expectation value (vev) -- a scenario we will refer to as monopole condensation.

\subsection{Symmetry Breaking and~$\t {S^3}$ Sigma Model from Monopole Vevs}

\label{introsquasheds3}

In this paper we will argue that symmetry breaking in massless~$N_f = 2$ QED$_3$ is due to the condensation of the~$q_m=1$ monopole in~\eqref{fundmono},
\begin{equation}\label{mvevintro}
    \langle \CM^i\rangle \neq 0~,
\end{equation}
which leads to the following symmetry-breaking pattern,
\begin{equation}\label{SSBintro}
    U(2) = {SU(2)_f \times U(1)_m \over \Z_2} \quad \longrightarrow \quad U(1)_\text{unbroken}~.
\end{equation}
Here~$U(1)_\text{unbroken}$ is the stabilizer group of the monopole vev~\eqref{mvevintro}, which we will discuss in more detail below.\footnote{~Group-theoretically, the breaking pattern~\eqref{SSBintro} is identical to the Higgsing pattern~$SU(2)_L \times U(1)_Y \to U(1)_\text{E\&M}$ due to the fundamental Higgs vev~$\langle h^i\rangle \neq 0$ in the standard model of particle physics.} In addition to~\eqref{SSBintro}, the vev~\eqref{mvevintro} also spontaneously breaks~$\CC$ and~$\CT$, but unbroken~$\t \CC$ and~$\t \CT$ symmetries can be constructed by mixing with the broken generators.

The symmetry-breaking pattern~\eqref{SSBintro} leads to three massless NGBs, described at low energies by the usual coset sigma model, which turns out to be a squashed three-sphere,
\begin{equation}\label{squasheds3}
{U(2) \over U(1)_\text{unbroken}} = SU(2) = \t{ S^3}~.
\end{equation}
Here we have used the notation~$\t S^3$ to indicate that the sphere metric is squashed in a~$U(2)$ symmetric fashion. This metric (and many other aspects of our story) are usefully described using Hopf coordinates, which arise by thinking of~$\t{ S^3}$ as a Hopf bundle (i.e.~an~$S^1$ fibration over a~$\C\P^1$ base), whose construction we now review.\footnote{~See for instance section 2.2 of~\cite{Seiberg:1996nz} for an introduction in a physically related context.} 

The monopole vev~\eqref{mvevintro} has non-vanishing~$U(2)$-invariant norm,
\begin{equation}
 |\langle \CM\rangle|^2 \equiv   \langle \b \CM_i\rangle \langle \CM^i\rangle > 0~.
\end{equation}
The~$U(2)$ orbit of the vev~\eqref{mvevintro} is precisely the squashed~$\t {S^3}$~in~\eqref{squasheds3}. Consider the following map from the monopoles~$\CM^i$ to a real unit vector field~$\vec n$,
\begin{equation}\label{hopfmapintro}
    \b \CM \, \vec \sigma \CM = |\langle \CM\rangle|^2 \vec n~, \qquad {\vec n}^2 = 1~.
\end{equation}
Note that~$\vec n$ transforms as an~$SU(2)_f$ triplet, but is neutral under~$U(1)_m$. The map from~$\CM^i$ to~$\vec n$ is the Hopf map, which exhibits~$\t {S^3}$ as a fiber bundle over the~$S^2$, or equivalently~$\C\P^1$, parametrized by~$\vec n$. For given~$\vec n$, the~$\CM^i$ in~\eqref{hopfmapintro} are unique up to an overall~$U(1)_m$ phase rotation, so that 
\begin{equation}\label{Mzetasigmaintro}
    \CM^i(\vec n, \sigma) = |\langle \CM\rangle| \zeta^i(\vec n) e^{i \sigma}~, \qquad \zeta^\dagger(\vec n) \vec \sigma \zeta(\vec n) = \vec n~, \qquad \sigma \sim \sigma + 2 \pi~. 
\end{equation}
Note that~$\sigma$ shifts under~$U(1)_m$ in such a way that~$e^{i\sigma}$ (and hence~$\CM^i$) has~$q_m = 1$. The~$U(2)$ invariant metric on~$\t {S^3}$ can now be written as follows,
\begin{equation}\label{HopfMetricIntro}
    ds^2(\t {S^3}) = r^2 d \vec n \cdot d \vec n + {e_0^2 \over 8 \pi^2} \left(d \sigma - \alpha\right)^2~.
\end{equation}
Here~$d \vec n \cdot d\vec n$ is the metric on a round~$S^2$ of unit radius, so that~$r$ is the radius of the base of the fibration; the one-form~$\alpha$ is a~$U(1)$ connection on the~$S^2$ base, whose curvature~$d \alpha / 2 \pi$ is the rotationally invariant unit area form~$\Omega$ on~$S^2$, $\int_{S^2} \Omega =1$. In other words, $\alpha$ is the connection of a unit Dirac monopole on~$S^2$. The angle~$\sigma$ parametrizes the~$S^1$ fiber over each point~$\vec n$ of the base; it has charge~$1$ under~$\alpha$ gauge transformations. The coefficient~$e_0$ determines the radius of the Hopf fiber; when the radii of base and fiber are related as~$4 r^2 = e^2_0/8 \pi^2$, the sphere is round and its isometry group is enhanced from~$U(2)$ to~$SO(4)$; as was already mentioned above, there is no reason to expect such accidental symmetry enhancement in the symmetry-breaking scenarios for massless QED$_3$ we consider here. On general grounds, we expect~$r^2$ and~$e_0^2$ to be of comparable magnitude; both should be~$\CO(1)$ when expressed in terms of the UV gauge coupling~$e^2$ in~\eqref{introQEDlag}, which sets the strong-coupling scale of the theory.

The Hopf coordinates provide a clean description of the stabilizer group~$U(1)_\text{unbroken}$ of the monopole vev~\eqref{mvevintro}. Given~$\langle \CM^i\rangle$, we can first determine the~$SU(2)_f$ triplet~$\vec n$ in~\eqref{hopfmapintro}. For simplicity, let us consider the north and south poles~$\vec n = \pm \vec e_3$ of the~$S^2$.\footnote{~Here we use~$\vec e_{1,2,3}$ to denote standard Cartesian unit vectors in~$SU(2)_f$ triplet space~$\R^3$.} These preserve the same flavor Cartan~$U(1)_f \subset SU(2)_f$, which we normalize so that~$\CM^1$ and~$\CM^2$ have~$U(1)_f$ charges~$q_f = 1$ and~$q_f = -1$, respectively. It follows from~\eqref{Mzetasigmaintro} that the corresponding monopole Hopf fibers are given by
\begin{equation}\label{monopolesNorthSouth}
    \CM^i(\vec e_3, \sigma) =  |\langle \CM\rangle|  e^{i\sigma} \begin{pmatrix}
        1 \\ 0
    \end{pmatrix} ~, \qquad   \CM^i(-\vec e_3, \sigma) = |\langle \CM\rangle|  e^{i\sigma} \begin{pmatrix}
        0 \\ 1
    \end{pmatrix}~.
\end{equation}
At the north pole, the stabilizer group that leaves~$\sigma$ invariant is thus\footnote{~Here we slightly abuse the notation and write linear combinations of~$U(1)$ symmetries to denote the corresponding relations between their charges. Since all~$U(2)$ representations have~$q_f \equiv q_m~(\text{mod}~2)$, it follows that the~$U(1)_\pm$ charges are integers.} 
\begin{equation}\label{unbrokennorth}
U(1)_\text{unbroken} = U(1)_- = \half \left(U(1)_m - U(1)_f\right) \quad \text{at north pole} \quad \vec n = \vec e_3~.
\end{equation}
The orthogonal linear combination~$U(1)_+ = \half\left( U(1)_m + U(1)_f\right)$ acts with charge~$+1$ on~$e^{i\sigma}$. At the south pole the roles of~$U(1)_\pm$ are reversed -- a hallmark of the fibration.

\subsection{Fermion Bilinears, Masses, and the Vafa-Witten Theorem}

\label{massesVWintro}

Since the monopole vev~\eqref{mvevintro} also induces a vev for the~$SU(2)_f$ triplet vector~$\vec n$ in~\eqref{hopfmapintro}, it is natural to ask whether the (non-monopole) fermion bilinear defined in~\eqref{TripletO}, which is also an~$SU(2)_f$ triplet, similarly acquires a vev. We will show below that this operator has the following effective description in the~$\t {S^3}$ sigma model at long distances,\footnote{~Note that~$\vec n$ is the only sigma-model operator without derivatives that has the same quantum numbers as~$\vec \CO$. The non-trivial statement is that the constant~$C$ must be strictly positive, and in particular cannot vanish. A similar phenomenon occurs for the chiral condensate in four-dimensional QCD, which (in standard four-dimensional conventions) must be negative when the quark mass is positive.}
\begin{equation}\label{OinIR}
\vec \CO = i \b \psi \vec \sigma \psi \qquad \xlongrightarrow{\text{RG flow}} \qquad C \vec n  + \left(\text{derivative terms}\right)~, \qquad C > 0~.
\end{equation}
Thus its vev is aligned with the~$U(1)_f \subset SU(2)_f$ Cartan already singled out by the monopole vev~\eqref{mvevintro}. If they were misaligned, this would spontaneously break the entire~$U(2)$ symmetry, a scenario that we will rule out momentarily using a variant of the Vafa-Witten theorem~\cite{Vafa:1983tf,Vafa:1984xh} that is suitably adapted to Abelian gauge theories with monopole operators.

Many arguments in this paper (including those in the spirit of Vafa and Witten) involve deforming the massless UV QED$_3$ theory via a real~$SU(2)_f$ triplet mass~$\vec m$ that couples to the fermion bilinear in~\eqref{TripletO} as follows,
\begin{equation}\label{tripletmassintro}
    \mathscr{L}_{\vec m} = \vec m \cdot \vec \CO = i \vec m \cdot \b \psi \vec \sigma \psi~, \qquad (\vec m)^* = \vec m~.
\end{equation}
On occasion, we will choose an explicit~$\vec m$ of the form
\begin{equation}\label{mzmass}
    \vec m = m \, \vec e_3~, \qquad m \in \R~,
\end{equation}
which explicitly breaks
\begin{equation}\label{u1u1intro}
  U(2) \qquad \xlongrightarrow{m \neq 0}\qquad {U(1)_f \times U(1)_m \over \Z_2}~.  
\end{equation}
It also preserves charge-conjugation~$\CC$, and the time-reversal symmetry~$\t \CT$ mentioned below~\eqref{SSBintro}.\footnote{~The definition of these symmetries requires a choice of~$SU(2)_f$ Cartan, because they involve a~$\pi$-rotation in~$SU(2)_f$ that flips the sign of that Cartan (see section~\ref{section: Lagrangian} for more detail). In the spontaneously broken case this Cartan is determined by the Hopf map~\eqref{hopfmapintro}.} 

If~$|\vec m| \ll e^2$ is sufficiently small, we can reliably analyze the mass deformation in the~$\t {S^3}$ sigma model description. Using~\eqref{OinIR}, we find that~\eqref{tripletmassintro} flows to
\begin{equation}\label{mdotnintro}
    \mathscr{L}_{\vec m} \qquad \xlongrightarrow{\text{RG flow}} \qquad C \vec m \cdot \vec n  + \left(\text{derivative terms}\right)~, \qquad C > 0~.
\end{equation}
Since the potential energy has an extra minus sign, this means that~$\vec n$ will precisely align with~$\vec m$. As is typical of spontaneous symmetry breaking, we can thus select different points on the~$\C\P^1$ base of the~$\t { S^3}$ by approaching~$\vec m = 0$ from different directions. Since the~$\vec n$ fluctuations acquire a mass thanks to~$\eqref{mdotnintro}$, we see from~\eqref{HopfMetricIntro} that we are only left with the compact massless scalar~$\sigma$ that parametrizes the Hopf fiber above the point~$\vec n \sim \vec m$.\footnote{~By contrast, explicitly adding a minimal~$q_m = 1$ monopole~$\CM^i$ to the Lagrangian of massless QED$_3$ leads to a single, trivially gapped vacuum in the~$\t {S^3}$ sigma model. This will be used in section~\ref{section: Anomalies}.} This in turn can be expressed (using standard Abelian duality in three dimensions) in terms of a free Maxwell field with gauge coupling~$e_0$ set by the radius of the Hopf fiber,
\begin{equation}\label{sigmaphotonintro}
- {e_0^2 \over 8 \pi^2 } d \sigma \wedge \star d \sigma + \cdots \quad \longleftrightarrow \quad - {1 \over 2 e_0^2} f \wedge \star f + \cdots~, 
\end{equation}
where the ellipses on both sides denote higher-derivative terms.

We are now in a position to comment on previously proposed symmetry-breaking scenarios for QED$_3$ in the literature. We will frame the discussion in terms of the Vafa-Witten theorems~\cite{Vafa:1983tf, Vafa:1984xh, Vafa:1984xg}; these apply to the theory deformed by a triplet mass~$\vec m = m \, \vec e_3$ as in~\eqref{mzmass}, which preserves the~$(U(1)_f \times U(1)_m )/\Z_2$, $\CC$, and~$\t \CT$ symmetries discussed around~\eqref{u1u1intro} (see also section~\ref{section:massiveSymmAnom}). As we explain in section~\ref{section: extending phase diagram}, the considerations of~\cite{Vafa:1983tf, Vafa:1984xh, Vafa:1984xg} lead to the following non-perturbative constraints:
\begin{itemize}
    \item[1.)] Time-reversal cannot be spontaneously broken.
    \item[2a.)] If no monopole operator condenses, then the entire~$(U(1)_f \times U(1)_m )/\Z_2$ symmetry is unbroken.
    \item[2b.)] If a monopole operator condenses, then~$(U(1)_f \times U(1)_m )/\Z_2$ is spontaneously broken to the~$U(1)$ stabilizer group of the monopole, which cannot be broken further. In other words, one linear combination of~$U(1)_f$ and~$U(1)_m$ is always unbroken. Note however that the ``vector-like'' $U(1)_f$ can be spontaneously broken, by mixing with~$U(1)_m$.\footnote{~This is a nice example in which the naive statement that vector-like fermion symmetries cannot be spontaneously broken is incorrect -- a possibility already emphasized in~\cite{Vafa:1983tf}.}
\end{itemize}
As usual, and following~\cite{Vafa:1983tf}, we expect the symmetries that are unbroken at~$\vec m \neq 0$ to remain unbroken as we take~$\vec m \to 0$. Our symmetry-breaking scenario is consistent with these constraints, and it realizes alternative~2b.) above. By contrast, any scenario that spontaneously breaks the entire~$U(2)$ symmetry, such as the hypothetical misalignment between the monopoles~$\CM^i$ and the fermion bilinear~$\vec \CO$ contemplated below~\eqref{OinIR}, is ruled out. 

The most common proposal in the literature, going back to~\cite{Pisarski:1984dj} (see~\cite{DiPietro:2015taa,Giombi:2015haa} for a more recent discussion with references) is that the fermion bilinear~$\vec \CO = i \b \psi \vec \sigma \psi$ gets a vev and spontaneously breaks $SU(2)_f \to U(1)_f$, leading to two NGBs described by a~$\C\P^1$ sigma model. In light of the constraints above, this proposal can be interpreted in two ways:
\begin{itemize}
    \item If no monopole condenses, then~$U(1)_m$ is unbroken and there are no additional NGBs -- and in particular no massless photon. (Recall from~\eqref{sigmaphotonintro} that a massless photon is (dual to) another NGB.) As we will explain below, anomaly matching implies that there must be additional dynamical degrees of freedom (which may be gapped or gapless) that are fibered over the~$\C\P^1$ sigma model. We discuss an example that matches all anomalies in section~\ref{section: TQFT deformation}; there the additional sector consists of a gapped, topological~$\Z_2$ gauge theory fibered over~$\C\P^1$.
\item If a monopole condenses, then~$U(1)_m$ is spontaneously broken, leading to exactly one more NGB~$\sigma$, or equivalently a massless photon (as in~\eqref{sigmaphotonintro}). The presence of a massless photon was already advocated in~\cite{Pisarski:1984dj},\footnote{~Roughly, this is because the vev of the fermion bilinear~$\vec \CO$ is also expected to induce a triplet mass~$\vec m \sim \langle \vec O\rangle$ for the fermions, leaving the~$\C\P^1$ and a massless photon at low energies. Precisely this scenario arises when we deform QED$_3$ in a particular symmetry-preserving fashion that we describe in section~\ref{section:deformationIntro}.} and with the benefit of hindsight we see that it should be interpreted in terms of monopole condensation, which (as already explained above) can in turn induce a suitably aligned vev for~$\vec \CO$. However, the presence of a massless photon in the IR does not uniquely determine which monopole condenses.\footnote{~Not all possibilities are compatible with anomaly matching, which requires the~$U(1)_f$ and~$U(1)_m$ charges~$q_f, q_m$ of the condensing monopole to both be odd (and thus non-zero).} We will argue that it is the minimal~$q_m = 1$ monopole~$\CM^i$ in~\eqref{fundmono}. Then the massless photon, or its dual~$\sigma$, is Hopf-fibered over~$\C\P^1$ and reconstitutes the~$\t {S^3}$ sigma model already described in section~\ref{introsquasheds3} above. 
\end{itemize}

\subsection{Phase Diagram of QED$_3$ with a Triplet Mass}

In section~\ref{section:massiveQED} we will consider the phase diagram of~$N_f = 2$ QED$_3$ as a function of the triplet mass~$\vec m$ in~\eqref{tripletmassintro}. We will first do this for~$\vec m \neq 0$, before taking~$\vec m \to 0$. This will allow us to argue for the symmetry-breaking pattern in section~\ref{introsquasheds3}, which is due to the vev of the monopole operator~$\CM^i$ in~\eqref{mvevintro}, with aligned triplet fermion bilinear~$\CO = i \b \psi \vec \sigma \psi$ in~\eqref{OinIR}. Without loss of generality, we choose the mass to be as in~\eqref{mzmass},
\begin{equation}
    \vec m = m \, \vec e_3~, \qquad m \in \R~,
\end{equation}
which preserves the~$(U(1)_f \times U(1)_m)/\Z_2$ symmetry in~\eqref{u1u1intro}.

In section~\ref{subsection:largemphase} we study the large-mass regime~$|m| \gg e^2$, where the fermions can be integrated out reliably at one-loop. This leads to a weakly-coupled Coulomb phase, described by free Maxwell theory (plus higher-derivative terms suppressed by~$|m|$), and importantly two Chern-Simons terms involving the background fields~$\widehat{A}_f, A_m$ for the unbroken~$U(1)_f$, $U(1)_m$ symmetries,\footnote{~Here~$e^2_m$ is the effective Maxwell gauge coupling as a function of the mass~$m$.}
\begin{equation}\label{Coulombphaseintro}
        \mathscr{L} = -{1 \over 2 e^2_m} \left|da\right|^2 + {q_f \over 2 \pi} \widehat{A}_f \wedge da + {q_m \over 2 \pi} A_m \wedge da + \cdots~. 
    \end{equation}
    Here the quantized Chern-Simons levels~$(q_f, q_m)$ determine the~$U(1)_f$, $U(1)_m$ charges of the minimal monopole operator~$e^{i \sigma}$ (expressed in terms of the dual photon~$\sigma$) that condenses. (The ellipsis denotes higher-order terms in~$da$.) In this weak-coupling regime, we inherit~$q_m = 1$ from the UV QED$_3$ theory, because the fermions do not carry~$U(1)_m$ charge; by contrast, integrating them out at one-loop gives~$q_f = \sign(m)$. No further corrections to the quantized levels are possible.
    
    We conclude that the monopole that condenses for large~$m > 0$ has exactly the same quantum numbers as the monopole at the north-pole of the~$\C\P^1$ in our symmetry-breaking scenario at~$m = 0$, see~\eqref{monopolesNorthSouth}; the monopole that condenses for large~$m < 0$ has the quantum numbers of the monopole at the south pole. Thus we see that the monopoles that condense at~$|m| \gg e^2$ are exactly the same monopoles that condense at~$m = 0$, if we extrapolate to the origin along rays in~$\vec m$ space. Indeed, the simplest and most economical scenario is that the large-$|m|$ Coulomb phase extends smoothly to all~$m \neq 0$.

Assuming this conjecture, and using anomaly matching, we argue in section~\ref{section:zeromass} that at~$m = 0$, all plausible alternatives to the squashed~$\t {S^3}$ symmetry-breaking scenario in section~\ref{introsquasheds3} can be dismissed. The only compelling alternative would be a~$U(2)$-invariant CFT, which (as we reviewed in section~\ref{secetion:qed3intro}) is implausible in light of recent bootstrap bounds. Note that the same conjecture also has a number of other implications,\footnote{~For instance, it implies (without referring to anomaly matching) that the theory at~$m= 0$ must be gapless, since a gapped theory would remain so for sufficiently small~$m$.} notably that the constant~$C$ in~\eqref{OinIR} ensuring a non-vanishing triplet vev~$\langle \vec \CO\rangle \neq 0$ aligned with the monopole vev~$\langle \CM^i\rangle \neq 0$, must in fact be positive, $C > 0$.\footnote{~Slightly more work is needed to argue that~$C$ has a strictly positive limit as~$\vec m \to 0$. This is shown around~\eqref{VW bound}.} 

In section~\ref{section: extending phase diagram} we give some arguments in favor of the conjecture above. In particular, we sketch a non-perturbative argument in the style of the Vafa-Witten theorem~\cite{Vafa:1983tf,Vafa:1984xh} that shows the exponential decoupling of all electrically charged degrees of freedom (whether fundamental or composite) at long distances, as long as~$m \neq 0$. This in turn implies a weakly-coupled massless photon NGB in the IR. We also explain the gaps that remain between this result and the full conjecture.

\subsection{Anomaly Matching}

\label{section:anomaliesintro}

\subsubsection{UV Anomalies and~$\theta =\pi$ in the~$\t {S^3}$ Sigma Model}

\label{section:thetaispiintro}

't Hooft anomaly matching for global symmetries provides a powerful constraint on all proposed IR scenarios. In particular, we will now show that anomaly matching requires the presence of a~$\theta$-term with coefficient~$\theta = \pi$ in the~$\t {S^3}$ sigma model described in section~\ref{introsquasheds3} resulting from the monopole vev~$\langle \CM^i\rangle$ in~\eqref{mvevintro}.

We briefly review the anomalies of~$N_f = 2$ QED$_3$ in section~\ref{section: Anomalies} (see also appendix~\ref{appcobcalcs}), where we confirm the results of~\cite{Benini:2017dus} (see also~\cite{Hsin:2024abm}) showing that the four-dimensional anomaly inflow action has path-integral weight
\begin{equation}\label{u2anomalyintro}
    \exp\left(i \pi \int_{\CM_4} c_2(U(2))\right)~.
\end{equation}
Here~$c_2(U(2))$ is the second Chern class of the background fields for the~$U(2)$ zero-form symmetry in~\eqref{u2symintro}. Note this is a mixed anomaly between~$U(2)$ and time-reversal~$\CT$, or indeed any orientation-reversing symmetry that pins the coefficient of~$c_2(U(2))$ to~$0$ or~$\pi$. The anomaly~\eqref{u2anomalyintro} shows that the theory cannot flow to a trivially gapped phase. In fact, such an anomaly cannot even be matched by a TQFT and requires gapless degrees of freedom in the IR \cite{Wang_2014, Wang_2016, Sodemann_2017,kantaroclaytodo}.\footnote{~An argument for this can be given using the theory of \cite{etingof2009fusioncategorieshomotopytheory}. The basic data of a $G$ action on a 3d TQFT is a permutation action $\rho$ on the anyons preserving the braiding, as well as fractionalization data in $H^2(BG,\cA^\rho)$, where $\cA$ is the group of abelian anyons, and this is twisted cohomology computed with $\rho$ action. $G = SU(2)$ is connected, so $\rho$ is trivial. It is also simply-connected, so $H^2(BSU(2),\cA) = 0$. So there is no way $SU(2)$ can have a non-trivial action on a 3d TQFT, in particular with any non-trivial anomaly. The $\pi c_2(U(2))$ anomaly meanwhile would imply a non-trivial $\pi c_2(SU(2))$ anomaly for the $SU(2)$ subgroup.} 

The~$\t {S^3}$ sigma model (with $f:\cM_3 \to S^3$) has a conventional~$\theta$-term that can be written in a local, gauge-invariant and~$U(2)$ symmetric fashion using the unit volume form~$\Omega_3$ on~$\t {S^3}$,
\begin{equation}\label{s3thetaintro}
    \exp\left(i \theta \int_{\CM_3} f^*\Omega_3\right)~, \qquad \theta \sim \theta + 2 \pi~.
\end{equation}
Only~$\theta = 0, \pi$ are compatible with time-reversal symmetry. Since the~$\theta$-term is~$U(2)$ invariant, we can couple it to~$U(2)$ background gauge fields. A straightforward calculation in equivariant cohomology (see appendix~\ref{appendixS3withU2action}) shows that, in the presence of background fields, $\Omega_3$ is extended to a well-defined three-form~$\t \Omega_3$, which satisfies~$d \t \Omega_3 = c_2(U(2))$. Thus extending~$\Omega_3$ in~\eqref{s3thetaintro} to~$\t \Omega_3$ in the presence of~$U(2)$ background fields leads to an arbitrary bulk~$\theta$-angle~$\exp(i\theta c_2(U(2)))$. Comparing with~\eqref{u2anomalyintro} then implies that we must choose~$\theta = \pi$. An alternative and instructive route to this conclusion will be explained below.

\subsubsection{Anomaly-Preserving Deformations of QED$_3$}
\label{section:deformationIntro}

Several aspects of our proposed symmetry-breaking scenario, driven by the monopole vev~\eqref{mvevintro} that is Hopf-fibered over the~$SU(2)_f$ triplet vev in~\eqref{OinIR}, are illuminated by engineering it as an explicit, weakly-coupled deformation of~QED$_3$ that preserves all symmetries and anomalies. An advantage of this approach is that anomaly matching is guaranteed, though checking this explicitly is not always straightforward and raises interesting questions in its own right.

To engineer this phase, we promote the triplet mass parameter~$\vec m$ in~\eqref{tripletmass} to a dynamical scalar field~$\vec \phi$ with exactly the same quantum numbers, and a canonical kinetic term, as well as a suitable scalar potential that preserves all symmetries. A very similar model -- with QED$_3$ in mind -- was considered in~\cite{Senthil:2005jk}, and more recently in~\cite{Dupuis:2019uhs}. Importantly, the Yukawa coupling~$\vec \phi \cdot \vec \CO$ that arises by promoting~$\vec m \to \vec \phi$ in the QED$_3$ mass term~\eqref{tripletmass} is automatically symmetric as well. Thus~$\vec \phi$ is a Hubbard-Stratonovich-like mean field for the fermion bilinear~$\vec \CO$; it allows us to consider weakly-coupled phases that are qualitatively similar to ones in which~$\vec \CO$ acquires a vev. The triplet field~$\vec \phi$ is also reminiscent of the scalar superpartner of the photon in versions of QED$_3$ with~$\CN=4$ supersymmetry, whose dynamics was analyzed in~\cite{Seiberg:1996nz}. Indeed, there are many parallels between our discussion here and the $\CN=4$ QED$_3$ theory with the smallest number of charged matter fields (i.e.~with a single~$\CN=4$ hypermultiplet of charge 1); these will be further explored in~\cite{wip}.

Let us dial the scalar potential for~$\vec \phi$ so that it gets a large vev~$|\langle \vec \phi \rangle| = v \gg e^2$. The radial mode of~$\vec \phi$ and the fermions acquire large masses and can be reliably integrated out. The vev~$\langle \vec\phi \rangle$ spontaneously breaks $SU(2)_f \to U(1)_f$, leading to a~$\C\P^1$ sigma model described by a unit vector field~$\vec n$ (so that~$\vec \phi = v \vec n$ at long distances). The only other massless particle at long distance is the photon, described by~$f = da$, with~$a$ the dynamical Spin$_c$ connection of the UV QED$_3$ theory. The low-energy Lagrangian after integrating out the massive modes takes the following form,
\begin{equation}\label{deformmodelintro}
    \mathscr{L}_\text{IR} = - {v^2 \over 2} \left |d \vec n\right|^2 - {1 \over 2 e^2} \left|da\right|^2 - a \wedge n^* \Omega_2 + \left(\text{higher derivatives}\right)~.
\end{equation}
Here the first two terms are the~$\C\P^1$ and Maxwell kinetic terms, while the third term is a Chern-Simons term that gauges the skyrmion current~$n^*\Omega_2$ of the~$\C\P^1$ model (i.e.~the pullback to spacetime of the unit area form~$\Omega_2$ on~$\C\P^1$) using the Spin$_c$ gauge field~$a$.\footnote{~This has the pleasing effect of trivializing the skyrmion symmetry of the~$\C\P^1$ model, which is not present in QED$_3$.} This one-loop exact Chern-Simons term has been computed explicitly by integrating out the fermions in the presence of the Yukawa coupling\cite{Abanov_2000,Hsin_2020,Choi_2022}; we will present an even simpler derivation in section~\ref{section: Deformation argument} by coupling to background fields.  

As was already emphasized in the supersymmetric context in~\cite{Seiberg:1996nz}, as well as in the context of QED$_3$ in~\cite{Senthil:2005jk}, the Chern-Simons term in~\eqref{deformmodelintro} has the effect of fibering the dual photon~$\sigma$ (see~\ref{sigmaphotonintro}) over the~$\C\P^1$ base, which leads to the squashed~$\t {S^3}$ sigma model with metric~\eqref{HopfMetricIntro}.\footnote{~Note that in our weakly-coupled model, the radius of the~$\C\P^1$ is large, $r \sim v$, while the radius of the Hopf fiber is set by the UV gauge coupling~$e^2$, and thus much smaller.} We review this in section~\ref{section: duality with S3}. 

A more subtle aspect of this story is that the Chern-Simons term in~\eqref{deformmodelintro} is not well-defined, because~$a$ is a Spin$_c$ connection. We carefully define it in section~\ref{section: Deformation argument}, where we also relate it to the discussion of the~$\C\P^1$ sigma model with Hopf term in~\cite{Freed_2018}. Indeed, we will show in section~\ref{section: duality with S3} that the properly defined, exponentiated Chern-Simons term gives rise to a sign~$(-1)^\text{Hopf Number}$ in the path integral.\footnote{~This is only precise if we take spacetime to be a sphere, $\CM_3 = S^3$. As discussed in~\cite{Freed_2018}, the Hopf term in the~$\C\P^1$ model requires a spin structure to be well-defined, because it turns the skyrmions into fermions. Here it can appear in a bosonic theory, because the skyrmion current of the model is Spin$_c$ gauged.} Upon dualizing~$a$ to the compact scalar~$\sigma$, this Hopf-number term gives rise to the~$\theta = \pi$ term in the~$\t {S^3}$ sigma model that we argued for in section~\ref{section:thetaispiintro} on the basis of anomaly matching. Indeed, we also explicitly check that~\eqref{deformmodelintro} (with properly defined Chern-Simons term) matches the anomaly~\eqref{u2anomalyintro}.

\subsection{Comments on~$N_f > 2$}

In section~\ref{section: Nf>2} we briefly describe a natural extension (consistent with all constraints) of our monopole-induced symmetry-breaking scenario to QED$_3$ with any even number~$N_f > 2$ of fermions. This is instructive, despite the fact that the bootstrap bounds reviewed in section~\ref{secetion:qed3intro} suggest that these theories in fact flow to symmetry-preserving CFTs. It was recently shown that precisely this scenario is relevant when we place QED$_3$ with any even~$N_f$ in a uniform magnetic field~\cite{Dumitrescu:2025vfp}.

\bigskip
\bigskip

\noindent \textit{Note Added: While this paper was being finalized, we became aware of~\cite{Chester:2024waw}, where symmetry breaking due to~$\langle \CM^i \rangle \neq 0$ is considered from a complementary point of view.}
\bigskip

\section{$N_f=2$ QED$_3$ in the UV: Symmetries and Anomalies}
\label{section: UV QED}
In this section, we study QED$_3$ with $N_f=2$ from the UV perspective. After a short review of its symmetries and of the quantum numbers of monopole operators, we determine the mixed 't Hooft anomaly between the $U(2)$ global symmetry and time reversal.

\subsection{Lagrangian, Monopoles, and Symmetries}
\label{section: Lagrangian}

We study QED$_3$ with $N_f = 2$ Dirac fermions. These are two-component complex spinors $\psi^i$ where $i$ is an $SU(2)_f$ flavor index. We give them unit gauge charge under a gauge field $a$, which must therefore be a Spin$^c$ connection.
The Lagrangian is
\begin{equation}
\mathscr{L}_{\rm QED} = - \frac{1}{2e^2} da \wedge \star da - i\Bar\psi_i \gamma^{\mu}(\partial_\mu-ia_\mu) \psi^i \,.   
\label{QEDLagrangian}
\end{equation}
Note that this theory does not need a background spin structure to be defined, and may thus be considered as a bosonic theory. Equivalently, it has no gauge-neutral fermion operators. The $\bZ_2$ center of $SU(2)_f$ acts on fundamental fermions $\psi^i$ as fermion number $(-1)^F$, which is equivalent to a gauge transformation with angle $\pi$, $\psi^i \rightarrow - \psi^i$. Thus, as far as operators constructed with fundamental fermions are concerned, the faithful global symmetry is only $SO(3)_f=SU(2)_f/\bZ_2$.

We now review monopole operators, which carry faithful $SU(2)_f$ representations.
Their quantum numbers can be determined by using the state-operator correspondence~\cite{Borokhov:2002ib}. The Hilbert space of zero modes of the free Dirac Hamiltonian on $\mathbb{R}_t \times S^2$, in a constant background of one unit of magnetic flux, $q_m = 1$, along $S^2$,
\begin{equation}
\int_{S^2} da = 2\pi \,,   
\end{equation}
is a Fock space of dimension $4$. Indeed, by the Atiyah-Singer theorem, each complex Dirac fermion $\psi^i$ contributes with two real zero modes, whose spin is $s=(|q_m|-1)/2=0$. Thus, there are $4$ degenerate spin-zero states for the free Dirac theory. In the case of QED, we need to impose the Gauss law constraint, which requires that the total gauge charge of physical states must be zero. This selects the two states which are created by acting with exactly one zero mode on the Fock vacuum, which transform as a doublet of $SU(2)_f$. We write the corresponding monopole operators as $\cM^i$, where $i$ is the $SU(2)_f$ doublet index.

We also assign magnetic $U(1)_m$ charge $q_m = 1$ to these $2\pi$-flux monopoles. The fact that non-monopole operators carry $SU(2)_f$ representations with integer spin implies that the faithful global symmetry is
\begin{equation}
    U(2) = \frac{SU(2)_f \times U(1)_m}{\bZ_2} \,,
\end{equation}
where the quotient identifies $-\bI_2 \in SU(2)_f$ with $-1 \in U(1)_m$.

The theory also enjoys discrete symmetries: a unitary charge-conjugation symmetry~$\cC$, and an anti-unitary time-reversal symmetry~$\cT$. These symmetries act as follows,\footnote{~The action on monopole operators can be determined by studying the zero-mode Fock space as in \cite{Cordova:2017kue}. Relative to that paper, note that their $\cT$ is our $\cC \cT$, and vice versa.}
\begin{equation}
\cC : \begin{cases}
    \psi^i \rightarrow (\psi^i)^* \\
    a_\mu \rightarrow -a_\mu \\
    \cM^i \rightarrow (\cM^i)^*
\end{cases}
\label{C on fermions}
\end{equation}
and\footnote{~Note that we are free to modify the action of~$\CT$ on monopoles~$\CM^i$ by a sign, by composing with~$(-1)^{q_m}$, which does not act on any other fields. Using this freedom, we choose~$\varepsilon_{ij} = \left(i \sigma^2\right)_{ij}$, so that~$\varepsilon_{12} = 1$.}
\begin{equation}
\cT : \begin{cases}
    \psi^i(t) \rightarrow \gamma^0 \psi^i(-t) \\
    a_0(t) \rightarrow a_0(-t) \\
    a_\mu(t) \rightarrow - a_\mu(-t) & \mu = 1,2 \\
    \cM^i(t) \rightarrow \varepsilon_{ij} (\cM^j)^*(-t)
\end{cases}
\label{T on fermions}
\end{equation}
These satisfy
\begin{equation}
\cC^2=1 \,, \qquad (\cC\cT)^2=\cT^2=(-1)^{q_m} \,.
\end{equation}
Notice that $\cC\cT$ commutes with $SU(2)_f$ transformations and anti-commutes with $U(1)_m$ transformations. 

Later, in section~\ref{section:massiveSymmAnom}, we will define~$\t \CT = \CT \CU_f$, where~$\CU_f = -i \sigma^2$ is an~$SU(2)_f$ transformation. This acts on the monopoles as follows,
\begin{equation}\label{tildeT}
    \t \CT : \CM^i \to (\CM^i)^*~, \qquad \CC \t \CT : \CM^i \to \CM^i~.
\end{equation}
Since both of these are anti-unitary, it follows that~$\t \CT$ preserves the monopole vev~$\langle \CM^i\rangle$, while~$\CC\t \CT$ complex-conjugates it. 

The most general mass term for the fermions may be written as
\begin{equation}
\mathscr{L}_{\rm mass} = i M^i_{~ j} \bar\psi_i \psi^j \,, \qquad \text{ with } M=M^\dagger \,.
\label{Lagrangian mass}
\end{equation}
We can decompose the matrix~$M$ into its~$SU(2)_f$ singlet and triplet parts, $M=m_0 \, \bI + \vec{m} \cdot \vec{\sigma}$, where $\vec \sigma$ are the Pauli matrices and both $m_0$ and $\vec{m}$ are real. Note that~$\vec m$ is precisely the triplet mass in~\eqref{tripletmassintro}. As a spurion, $M$ transforms as follows, 
\begin{equation}
\begin{split}
SU(2)_f &: M \rightarrow U^\dagger M U \,, \\
\cC &: M \rightarrow M^t \,, \\
\cT &: M \rightarrow -M^t \,.
\label{mass transformations}
\end{split}
\end{equation}
Notice that the massless point $M=0$ in \eqref{QEDLagrangian} is enforced by $\cC \cT$ symmetry alone.

\subsection{Anomalies}
\label{section: Anomalies}
Let us now discuss the anomalies of the theory. We couple the Lagrangian \eqref{QEDLagrangian} to a $U(2)$ background field $\cA$, which we decompose into an $SO(3)_f$ gauge field $A_f^I$ and a $U(1)_m$ gauge field $A_m$, related by
\begin{equation}
    \cA = A_m \, \mathbb{I}_2 + A_f^I \, \frac{\sigma^I}{2}~,
\end{equation}
where $\sigma^I/2$ are the generators of the $\mathfrak{su}(2)_f$ Lie algebra, with $I=1,2,3$ an adjoint index. Importantly, $dA_m/2\pi$ may have half-integer periods, satisfying
\begin{equation}\label{w2so(3)f&dAm}
\frac{1}{2} \oint_{\Sigma_2} w_2(SO(3)_f) = \oint_{\Sigma_2} \frac{dA_m}{2\pi} \text{ mod }\bZ \,,
\end{equation}
on closed surfaces $\Sigma_2$. This encodes the statement that even-charge monopoles transform in integer-spin representations of~$SU(2)_f$, while odd-charge monopoles transform in half-integer spin~$SU(2)_f$ representations. 

We can write the Lagrangian with this background as
\begin{equation}
\mathscr{L}_{\rm QED} = - \frac{1}{4e^2} f^{\mu\nu} f_{\mu\nu} - i\Bar\psi_i \gamma^{\mu}\left((\partial_\mu-ia_\mu)\bI_2-i(A_f)^I_\mu \frac{\sigma^I}{2} \right)^i_{~ j} \psi^j + \frac{1}{2\pi} da \wedge A_m \,.
\label{QEDlagrangianbackground}
\end{equation}
When we turn on background fields for the $U(2)$ symmetry, the $U(1)_g$ gauge symmetry becomes a $\bZ_2$ extension of both $U(2)$ and the Lorentz group,
\begin{equation}
\frac{U(1)_g \times U(1)_m \times SU(2)_f \times {\rm Spin(2,1)}}{\bZ_2 \times \bZ_2} \,.    
\end{equation}
One $\mathbb{Z}_2$ quotient identifies the center of $SU(2)_f$ and the element $-1$ of $U(1)_m$, leading to the relation \eqref{w2so(3)f&dAm}, namely to the fact that the global symmetry is $U(2)$. The other $\mathbb{Z}_2$ quotient modifies the quantization of the Spin$^c$ connection to\footnote{~This expression is just a mnemonic, since all oriented 3-manifolds are spin and so the integrals of $w_2(T\cM_3)$ are always even. However, it reminds us that diffeomorphisms induce non-trivial gauge transformations of $a$ since it is a Spin$^c$ connection. It also plays a crucial role in defining Chern-Simons terms for $a$, as not all 4-manifolds are spin.}
\begin{equation}
\oint_{\Sigma_2} \frac{da}{2\pi} = \frac{1}{2} \oint_{\Sigma_2} w_2(T\cM_3) +  \frac{1}{2}\oint_{\Sigma_2} w_2(SO(3)_f) \quad \text{ mod } \bZ \,.
\label{correctspinc}
\end{equation}
Notice that the second term on the right-hand side can also be written in terms of the first Chern class $c_1$ of $U(2)$, by using \eqref{w2so(3)f&dAm} and the fact that
\begin{equation}
\frac{1}{2} \oint_{\Sigma_2} c_1(U(2)) \equiv \frac{1}{2} \oint_{\Sigma_2} \frac{\tr \cF}{2\pi} = \oint_{\Sigma_2} \frac{dA_m}{2\pi} \,.
\label{c1U(2)}
\end{equation}

We will now compute the anomaly. For simplicity, we focus on the $U(1)_f$ Cartan subgroup of $SU(2)_f$. We define its background field by
\begin{equation}
    \hat A_f = \frac{1}{2} A_f^{I=3}.
\end{equation}
With this definition, the two fermions decouple, as $\psi^1$ has charge 1 under $a + \hat A_f$ and $\psi^2$ has charge 1 under $a - \hat A_f$.
We give a classification of all possible $U(2)$, $\cC$, and $\cT$ anomalies in Appendix \ref{appendixU2result}, and this analysis will turn out to be sufficient to determine the full anomaly.

The theory of a single 3d Dirac fermion with charge 1 under a gauge field $A$ suffers from the well-known parity anomaly~\cite{Redlich:1983dv, Redlich:1983kn, Niemi:1983rq, Alvarez-Gaume:1983ihn}, which may be cancelled by the four-dimensional action 
\begin{equation}
\pm \pi I_1[A] = \pm \left( \frac{\pi}{8}\sigma + \frac{\pi}{2} \int_{\cM_4} \frac{dA}{2\pi} \wedge \frac{dA}{2\pi} \right)\,,
\label{4d anomaly}
\end{equation}
where the choice of the overall sign is arbitrary (it depends on the choice of regularization scheme), and $\sigma$ is the signature of~$\cM_4$; it is a multiple of $16$ for spin manifolds. In our case, we have two Dirac fermions, one coupled to $a+\Hat{A}_f$ and the other coupled to $a-\Hat{A}_f$. For our purposes it is convenient (and sufficient) to regularize the fermions with opposite signs in \eqref{4d anomaly}. This preserves a certain definition of time-reversal, but breaks the~$U(2)$ symmetry, leading to
\begin{equation}
\pi \left( I_1[a+\Hat{A}_f] - I_1[a-\Hat{A}_f] \right) = \frac{1}{2\pi} \int_{\cM_4} da \wedge d\Hat{A}_f \,.
\label{fermion regularization}
\end{equation}
When we combine this term with the coupling to the~$U(1)_m$ background field~$A_m$ in \eqref{QEDlagrangianbackground}, we find
\begin{equation}
\begin{split}
    S_{\rm bulk}[A_m,\Hat{A}_f] &= \frac{1}{2\pi} \int_{\cM_4}  da \wedge (dA_m \pm d\Hat{A}_f) 
    \\ &= \pi \int_{\cM_4} \left( {d A_m \over 2\pi} \wedge {d A_m \over 2\pi} - {d \Hat A_f \over 2\pi} \wedge {d \Hat A_f \over 2\pi} \right) 
    = \pi \int_{\cM_4} \frac{dA_+}{2\pi} \wedge \frac{dA_-}{2\pi} \,.
\end{split}
\label{anomalyAmAf}
\end{equation}
Here we have used~\eqref{correctspinc} and the Wu formula, and we have defined the conventional~$U(1)$ connections
\begin{equation}\label{apmdef}
A_+ \equiv A_m + \Hat{A}_f  \,, \qquad A_- \equiv A_m - \Hat{A}_f \,.  
\end{equation}
By the Whitney sum formula, the anomaly in~\eqref{anomalyAmAf} corresponds uniquely to the following $U(2)$ form,
\begin{equation}
S_{\rm bulk}[\cA] = \pi \int_{\cM_4} c_2(U(2)) = \frac{1}{8\pi} \int_{\cM_4} \bigg( \tr\cF \wedge \tr\cF - \tr\left( \cF \wedge \cF \right) \bigg) \,, 
\label{anomaly as c2}
\end{equation}
where $\cF$ is the curvature of the $U(2)$ gauge field $\cA$. Note that~\eqref{anomaly as c2} agrees with the results of \cite{Benini:2017dus} (see also \cite{Hsin:2024abm}), where a~$U(2)$-preserving, but time-reversal breaking regulator was used. Indeed, \eqref{anomaly as c2} is a parity anomaly, i.e.~a mixed anomaly between~$U(2)$ and an orientation-reversing symmetry that pins the coefficient of~$c_2(U(2))$ to~$0$ or~$\pi$. There is no pure $U(2)$ anomaly, as was already shown in \cite{Benini:2017dus}.

In principle, there may be other possible anomalies involving the discrete symmetries $\cC$ and $\cT$. According to our classification in Appendix \ref{appendixU2result}, we have the following options, none of which end up being realized in~$N_f = 2$ QED$_3$. Here we give short arguments ruling them out in turn:

\begin{itemize}

\item A pure $\cT$ anomaly, which would amount to a gravitational theta term with $\theta = \pi$,
\begin{equation}
\pi \int_{\cM_4} w_2(T\cM_4) \cup w_2(T\cM_4) \,.
\end{equation}
This would already be visible from the calculation above, but it does not appear given that there is a choice of regulator which preserves time-reversal invariance, and the spacetime contribution encoded in $\sigma$ cancels in \eqref{fermion regularization}. Thus, this anomaly is not present.

\item A pure $\cC$ anomaly, which would amount to a theta term with $\theta=\pi$ for the background $\bZ^\cC_2$ gauge field $A_\cC$ (such that $[dA_\cC] = 0$ mod $2$),
\begin{equation}
\pi \int_{\cM_4} \frac{dA_\cC}{2} \cup \frac{dA_\cC}{2} \,.
\end{equation}
This anomaly can be ruled out by deforming the theory with the $U(2)$-preserving and $\cT$-breaking mass deformation $\mathscr{L}_m=im_0\Bar\psi_i\psi^i$. Integrating out the fermions for large $|m_0|$, we get a pure Chern-Simons theory $U(1)_{\pm 1}$ (the sign of the level is given by the sign of $m_0$), which is an invertible theory\footnote{~In particular, the Hilbert space of the theory quantized on any Riemann surface consists of a single state and the partition function is just a phase.} and has a unique trivially gapped vacuum (see Appendix \ref{spincchernsimonsappendix}).

\item An anomaly mixing $c_1(U(2))$ with $\cC$, which would amount to a mixed theta term with $\theta=\pi$,
\begin{equation}
\pi \int_{\cM_4} c_1(U(2)) \cup \frac{dA_\cC}{2} \,.
\label{mixanc1C}
\end{equation}
This can also be ruled out by a deformation argument, albeit a slightly more involved one. First, we turn on a large mass deformation $\mathscr{L}_m=im(\Bar\psi_1\psi^1-\Bar\psi_2\psi^2)$ for the fermions, which preserves $\cC$ and the subgroup \eqref{u1u1intro} of $U(2)$. As we will show in section~\ref{subsection:largemphase}, the resulting theory is a weakly-coupled Coulomb phase described by nearly free Maxwell theory, or equivalently the dual photon~$\sigma$. Then the condensing monopole~$e^{i \sigma}$ has charge $(\pm 1,1)$ under $(U(1)_f \times U(1)_m)/\bZ_2$, where the sign of the $U(1)_f$ charge is given by the sign of $m$. This corresponds to have charge $(1,0)$ under $(A_\pm,A_\mp)$, where $A_\pm = A_m \pm \Hat{A}_f$.
Then, we add the deformation $\Delta\mathscr{L}=\cos \sigma$, which further breaks the global symmetry to $U(1)_\mp \rtimes \bZ_2^\cC$ and yields a trivially gapped vacuum. As we show in Appendix \ref{appendixU2result}, this unbroken subgroup would inherit the mixed anomaly \eqref{mixanc1C} as
\begin{equation}
\pi \int_{\cM_4} \frac{dA_\mp}{2\pi} \cup \frac{dA_\cC}{2} \,,
\end{equation}
which cannot be matched by a trivially gapped vacuum. Thus, this anomaly needs to vanish as well.

\end{itemize}

\section{Phase Diagram of~$N_f = 2$ QED$_3$ with~$SU(2)_f$ Triplet Mass~$\vec m$}

\label{section:massiveQED}

In this section we will explore the phase diagram of~$N_f=2$ QED$_3$ in the presence of an arbitrary real triplet mass~$\vec m$, introduced in~\eqref{tripletmassintro}, which we repeat here, 
\begin{equation}\label{tripletmass}
    \mathscr{L}_{\vec m} = \vec m \cdot \vec \CO = i \vec m \cdot \b \psi \vec \sigma \psi~, \qquad (\vec m)^* = \vec m~.
\end{equation}
In particular, we will do so through the lens of~\cite{Vafa:1983tf, Vafa:1984xh, Vafa:1984xg}. 

\subsection{Residual Symmetries and Anomalies}

\label{section:massiveSymmAnom}

Since~$\vec m$ transforms in the triplet representation of~$SU(2)_f$, it suffices to fix a particular direction, which we choose as in~\eqref{mzmass},
\begin{equation}\label{zmassdef}
\vec m = m \, \vec e_3~, \qquad m \in \R~, \qquad \mathscr{L}_{m} = i m \left(\bar \psi_1 \psi^1 - \bar \psi_2 \psi^2\right)~.
\end{equation}
In principle we could further restrict~$m > 0$, but it will be instructive to consider both signs for~$m$. Thus, we will first study the phase diagram as a function of~$m$, and then contemplate the consequences of covariantizing with respect to~$SU(2)_f$. 

Let us summarize the symmetries of the mass-deformed theory:
\begin{itemize}
    \item The~$U(2)$ symmetry is explicitly broken to its~$( U(1)_f \times U(1)_m)/  \bZ_2$ Cartan. We will denote the~$U(1)_f$ and~$U(1)_m$ charges by~$q_f, q_m \in \Z$. By virtue of the~$\Z_2$ quotient that enforces the action of the~$U(2)$ Weyl group, they satisfy~$q_f \equiv q_m ~ (\text{mod } 2)$. Thus the monopole operators~$\CM^i$ have charges
    \begin{equation}\label{monopoleqfqm}
        q_m(\CM^i) = 1~, \qquad q_f(\CM^1) = 1~, \qquad q_f (\CM^2) = -1~.
    \end{equation}
    
    \item It follows from~\eqref{mass transformations} that
    \begin{equation}\label{unbrokenctt}
        \cC: m \to m~, \qquad \cT: m \to -m~.
    \end{equation}
    Thus charge conjugation~$\cC$ is preserved, while time reversal can be combined with a broken~$SU(2)_f$ rotation~$\cU_f = - i \sigma^2 $ to obtain the following unbroken time-reversal symmetry,\footnote{~Note that there is a sign ambiguity associated with the central element of~$SU(2)_f$ (which does not act on~$\vec m$), which is nothing but~$(-1)^{q_m} \in U(1)_m$. We choose the sign of~$\CU_f$ so that~$\t \CT$ will be unbroken in the phases we encounter. Since these will be Coulomb phases, where~$(-1)^{q_m}$ is spontaneously broken, only one choice~$\t \CT$ will be unbroken.}
    \begin{equation}\label{Ttildedef}
        \widetilde \cT = \cT \cdot \cU_f~, \qquad \cU_f = - i \sigma^2 ~. 
    \end{equation}
    Using~\eqref{C on fermions} and~\eqref{T on fermions}, we have the following actions on the monopoles,
    \begin{equation}\label{CtTonMonopole}
        \CC : \CM^i \to (\CM^i)^*~, \qquad \t \CT : \CM^i \to (\CM^i)^*~.
    \end{equation}
    Note that~$\t \CT$, being anti-unitary, preserves the $c$-number monopole vevs~$\langle \CM^i\rangle$, while the unitary~$\CC$ is only preserved if these vevs are real. 
\end{itemize}

As was already explained in section~\ref{section: Anomalies}, the anomalies remain non-trivial at non-zero~$m$, where they take the form in~\eqref{anomalyAmAf}, which we repeat here,
\begin{equation}\label{U(1)anomalyii}
S_{\rm bulk}[A_m,\Hat{A}_f] = \pi \int_{\cM_4} \left( {d A_m \over 2\pi} \wedge {d A_m \over 2\pi} - {d \Hat A_f \over 2\pi} \wedge {d \Hat A_f \over 2\pi} \right)~.
\end{equation}
Note that the unbroken~$\widetilde \cT$ and~$\CC\t \CT$ symmetries do indeed pin the anomaly coefficient in~\eqref{U(1)anomalyii} to~$0$ or~$\pi$, i.e.~the anomaly remains non-trivial. In particular, this means that we cannot find a trivially gapped phase for any~$m$.

\subsection{The Large~$|\vec m|$ Coulomb Phase}
\label{subsection:largemphase}

When~$|m| \gg e^2$ is much larger than the strong-coupling scale of the theory, set by the UV gauge coupling~$e^2$, we can reliably integrate out the fermions. At low energies, the resulting phase is described by free Maxwell theory plus higher-derivative corrections of Euler-Heisenberg type, i.e.~it is a Coulomb phase. In particular, the~$U(1)_m$ magnetic symmetry is spontaneously broken, with the photon being the associated NGB. The photon can be dualized to a compact scalar~$\sigma \sim \sigma + 2 \pi$, and the fundamental monopole operator~$e^{i \sigma}$ of Maxwell theory is the symmetry-breaking order parameter. 

We would like to understand which~$U(1)_m$-charged monopole operator of the UV QED theory flows to~$e^{i\sigma}$, and thus triggers the symmetry breaking. The~$U(1)_f$ and~$U(1)_m$ charges $(q_f, q_m)$ of the monopole operator~$e^{i\sigma}$ are determined by mixed Chern-Simons terms in the low-energy effective action~\eqref{Coulombphaseintro} for the Maxwell field in the deep IR, which we repeat here
\begin{equation}\label{qfqmCSterms}
 \mathscr{L} = -{1 \over 2 e^2_m} \left|da\right|^2 + {q_f \over 2 \pi} \Hat{A}_f \wedge da + {q_m \over 2 \pi} A_m \wedge da + \cdots~. 
\end{equation}
Here~$e_m^2$ is the effective Maxwell gauge coupling as a function of~$|m|$, which approaches the UV gauge coupling~$e^2$ at large~$|m|$.\footnote{~See~\eqref{ir limit propagator} (with~$N_f = 2$) for the one-loop corrected~$e_m^2$.} By retracing the steps below~\eqref{fermion regularization}, it can be shown that the anomaly~\eqref{U(1)anomalyii} is matched if both $q_f$ and $q_m$ are odd, consistent with the~$U(2)$ selection rule~$q_f \equiv q_m~(\text{mod } 2)$ discussed above~\eqref{monopoleqfqm}. For instance, in the weakly-coupled large-$|m|$ regime, we find\footnote{~The fermions do not carry~$U(1)_m$ charge, so that~$q_m = 1$ just follows from the UV Lagrangian~\eqref{QEDlagrangianbackground}. By contrast, $\psi^{i = 1}$ has~$q_f = 1$ and mass~$m$, while~$\psi^{i = 2}$ has~$q_f = -1$ and mass~$-m$. Since both have gauge charge~$+1$, it follows that each of them contributes~$+{1 \over 2} \sign(m)$ to~$q_f$ in \eqref{qfqmCSterms}, leading to~\eqref{qfqmCS}. }
\begin{equation}\label{qfqmCS}
     q_f = \sign(m)~, \qquad q_m = 1~, \qquad |m| \gg e^2~.
\end{equation}

From~\eqref{qfqmCS}, we can unambiguously conclude that the UV monopole operator that condenses is~$\cM^{i = 1}$ when~$m > 0$ and~$\cM^{i = 2}$ when~$m < 0$,
\begin{equation}\label{mzmonovev}
\langle \cM^i\rangle = \langle \CM \rangle \left( \theta (m) \delta^{i 1} + \theta(-m) \delta^{i2}\right)~, \qquad \langle \CM \rangle \in \bC^*.
\end{equation}
It follows that, for any sign of the mass, the pattern of spontaneous symmetry breaking is
\begin{equation}
    { U(1)_f \times U(1)_m \over \bZ_2} \qquad \longrightarrow \qquad U(1)_\text{unbroken} \,.
\end{equation}
In either case there is one NGB (the photon), but the unbroken symmetry group depends on the sign of~$m$,
\begin{equation}\label{unbrokenqfqm1}
    U(1)_\text{unbroken} = \begin{cases}
			U(1)_- \,  & \text{ if } m > 0 \,, \\
            U(1)_+ \,  & \text{ if } m < 0 \,.
		 \end{cases}
\end{equation}
Here~$U(1)_\pm = \half (U(1)_m \pm U(1)_f)$ are the symmetries we have already encountered in section~\ref{introsquasheds3}, around~\eqref{unbrokennorth}, when discussing the non-trivially fibered unbroken symmetries in the~$\t {S^3}$ sigma model; here they are similarly fibered over the~$S^2$ of~$\vec m$-directions at large~$|\vec m|$. Note that the~$U(1)_\pm$ background gauge fields~$A_\pm = A_m \pm A_f$ are standard~$U(1)$ connections. 

Let us discuss the action of discrete symmetries~$\CC$ and~$\t \CT$ on the monopole vevs~\eqref{mzmonovev}. It follows from~\eqref{CtTonMonopole} that~$\t \CT$ is unbroken for any complex~$\langle \CM\rangle \in \C^*$, while unbroken~$\CC$ requires~$\langle \CM\rangle$ to be real. This can always be achieved using one of the broken generators to shift~$\sigma$, so that there is an unbroken charge-conjugation symmetry in any vacuum.

Recalling the~$SU(2)_f$-covariant description~\eqref{Mzetasigmaintro} of the symmetry-breaking pattern~\eqref{monopolesNorthSouth}, we can also covariantize~\eqref{mzmonovev}:  for any~$\vec m \neq 0$, the monopole operator that is proportional to~$e^{i\sigma}$ in Maxwell theory is given by\footnote{~To recover~\eqref{mzmonovev}, note that~$m^1 = m^2 = 0$ implies that~$(\zeta^1)^* \zeta^2 = 0$. Together with the constraint that~$m^3 = m$ and~$|\zeta^1|^2 - |\zeta^2|^2$ must have the same sign, this shows that only~$\zeta^1$ is nonzero when~$m > 0$ and only~$\zeta^2$ is nonzero when~$m < 0$.}
\begin{equation}\label{zetaspinor}
    \cM^i = |\langle \CM\rangle| \zeta^i(\hat m) e^{i \sigma}~, \qquad \zeta^\dagger(\hat m) \vec \sigma \zeta(\hat m) = \hat m \equiv \vec m/ |\vec m|~.
\end{equation}
As in~\eqref{Mzetasigmaintro}, the undetermined phase is accounted for by shifts of the dual photon~$\sigma$. This shows that~$\sigma$ is precisely the fiber of the Hopf map~$\cM^i \to \vec m$, so that both the broken and the unbroken~$U(1)$ symmetries are non-trivially fibered over space of~$\vec m$-directions specificed by the unit vector~$\hat m$.

\subsection{Extending the Phase Diagram to all~$\vec m \neq 0$}
\label{section: extending phase diagram}
The simplest, most economical conjecture, is that the Coulomb phase at large~$|\vec m| \gg e^2$ described above in fact extends to all~$\vec m \neq 0$. The reader who is willing to accept this conjecture is directed to section~\ref{section:zeromass}, where we will use it to constrain the physics at the origin~$\vec m = 0$.

Since the theory is strongly coupled when $|\vec m| \lesssim e^2$, this conjecture is by no means obvious. Indeed, experience shows that three-dimensional gauge theories with suitable matter (and possibly also Chern-Simons terms for the dynamical gauge fields) often realize quantum phases that cannot be reached from any conventional weak-coupling regime, see e.g.~\cite{Komargodski:2017keh,Gomis:2017ixy,Choi:2018ohn,Choi:2018tuh,Cordova:2018qvg,Argurio:2019tvw,Armoni:2019lgb,Aitken:2019mtq,Choi:2019eyl} for an incomplete list. 

In this section we will give some evidence against the existence of such a quantum phase in QED$_3$ with $N_f =2$ flavors for any non-vanishing triplet mass~$\vec m \neq 0$. In particular, we will argue for a bound on all electrically charged matter (elementary or composite) in the spirit of the Vafa-Witten theorems~\cite{Vafa:1983tf, Vafa:1984xh, Vafa:1984xg}, and in turn for the existence of a weakly-coupled massless photon in the IR. 

We begin by reviewing what the results of~\cite{Vafa:1983tf, Vafa:1984xh, Vafa:1984xg} tell us about QED$_3$, in part to emphasize the somewhat unusual symmetry-breaking patterns that can arise due to monopole operators, which mix the~$U(1)_m$ magnetic symmetry and flavor symmetries. This possibility, already noted in~\cite{Vafa:1983tf, Vafa:1984xh, Vafa:1984xg} and even earlier in~\cite{Affleck:1982as}, is closely related to the Chern-Simons terms~\eqref{qfqmCSterms} that we already encountered in the large-$|\vec m|$ Coulomb phase.

\subsubsection{The Vafa-Witten Theorems}
\label{section: Vafa-Witten}

The Vafa-Witten theorems~\cite{Vafa:1983tf, Vafa:1984xh, Vafa:1984xg} impose non-perturbative restrictions on vector-like gauge theories (without Chern-Simons terms), whose Euclidean path-integral measure is positive definite after turning on suitable Dirac masses compatible with time reversal. In the context of~$N_f = 2$ flavor QED$_3$, Vafa and Witten considered the mass deformation~\eqref{zmassdef}, for fixed positive~$m$,\footnote{~Of course any fixed ray in triplet-mass space~$\vec m \in \R^3 - \{0\}$ can be analyzed in this way.}
\begin{equation}\label{vwmass}
    \mathscr{L}_m = i m \left(\b \psi_1 \psi^1 - \b \psi_2 \psi^2\right)~, \qquad m > 0~.
\end{equation}
This theory can be regulated in such a way that the Euclidean measure is indeed positive definite. 

\paragraph{Unbroken Time Reversal:} 
One consequence of measure-positivity, advocated in~\cite{Vafa:1984xg}, is that there is an unbroken time-reversal symmetry. As discussed around~\eqref{U(1)anomalyii}, this ensures that the anomaly must be matched by non-trivial IR degrees of freedom charged under the $(U(1)_f \times U(1)_m)/\Z_2$ symmetry. In the Coulomb phase we explored in section~\ref{subsection:largemphase} we saw that the~$\t \CT$ symmetry preserved by the mass term is unbroken. The anomaly is matched by the free Maxwell theory in the IR, but in principle other scenarios are compatible with anomaly matching (see section~\ref{section: IR QED}).

\paragraph{Unbroken Flavor Symmetries:} In~\cite{Vafa:1983tf}, Vafa and Witten used measure-positivity at finite~$m$ to obtain bounds on vector-like current correlators. Roughly speaking, they found a bound on a suitably smeared, gauge-invariant version (the technical details of which will not be important here) of the Dirac propagator $S_a(x,y)$ in an arbitrary fixed background~$a$ for the dynamical gauge fields. Schematically,
\begin{equation}\label{vwpropbound}
    |S_a(x, y) | \lesssim e^{- m |x - y|}~.
\end{equation}
Here~$m>0$ is the bare mass term~\eqref{vwmass} in the Lagrangian. Note that all arguments are carried out in a theory with a (suitably gauge-invariant) UV cutoff, which is not spelled out explicitly. Given any vector-like current~$J^I_\mu \sim \bar \psi \gamma_\mu T^I \psi$ (with~$T^I$ a suitable generator of the flavor symmetry Lie algebra), we can contract the fermions, which leads to two propagators, each of which is bounded as in~\eqref{vwpropbound}. Averaging over the positive measure then leads to a bound of the schematic form
\begin{equation}
\braket{J_\mu^I(x) J_\nu^K(y)} \lesssim e^{-2m|x-y|}~.
\label{VW bound}
\end{equation}
Since this decays exponentially in position space, the current cannot create a massless NGB from the vacuum (which would lead to power-law decay in position space, or a single-particle massless pole in momentum space), and hence the symmetry is not spontaneously broken.

Note that the bound~\eqref{VW bound} holds as long as we only contract fermions at distinct spacetime points~$x$ and~$y$. These ``connected diagrams'' consist of a single fermion loop in a bosonic background that has not yet been path-integrated over (see figure 3a in~\cite{Vafa:1983tf}).\footnote{~This is not the same as the (strictly perturbative) notion of a sum over all connected Feynman diagrams in the full theory, with dynamical fermions and photons.} This is inescapable as long as the current~$J_\mu^I$ is charged under a global symmetry that is only carried by fermions, but not the gauge bosons~$a$ (or other bosonic fields that may be present) over which we subsequently path-integrate. 

It is interesting to apply this bound to the~$SU(2)_f$ currents~$J^{1}_\mu \mp i J_\mu^2$, which carry~$U(1)_f$ charges~$\pm 2$. It implies that they create a particle of mass~$\gtrsim m$. Note that these currents are explicitly broken when~$m \neq 0$, and -- in the scenario of section~\ref{introsquasheds3} -- spontaneously broken at~$m = 0$, where they create the two NGBs parametrizing the $\mathbb{CP}^1$ base~$\vec n$ of the squashed~${\t S}^3$ sigma model. Comparing with~\eqref{mdotnintro}, we see that the coefficient~$C$ that appears there must be non-zero in the limit~$\vec m \to 0$. 

If instead we consider a current that is neutral under all global symmetries that are only carried by fermions, we must also consider fermion contractions at the same spacetime point. This leads to ``disconnected diagrams'' consisting of two fermion loops, with one current attached to each loop (see figure 3b in~\cite{Vafa:1983tf}). These need not exponentially decay at long distances. Precisely this important loophole can in principle arise in QED$_3$, as already emphasized in \cite{Vafa:1983tf, Vafa:1984xh}. We now review this phenomenon, while adding some comments related to the Chern-Simons terms~\eqref{qfqmCSterms} along the way.

\paragraph{Possible Mixing of~$U(1)_f$ and~$U(1)_m$ Symmetries:} 

Consider the current~$J_\mu^f$ of the~$U(1)_f$ flavor symmetry that is present at nonzero fermion mass~$m$,
\begin{equation}
J^f_\mu = \Bar\psi_i \gamma_\mu (\sigma^3)^{i}_{~ j} \psi^j \,.
\end{equation}
This current assigns charge~$+1$ to~$\psi^1$ and~$-1$ to~$\psi^2$, so that it is exactly the current that couples to~$\hat A_f = \half A_f^{I = 3}$ in~\eqref{QEDlagrangianbackground}, and that also appears in the Coulomb phase Chern-Simons terms~\eqref{qfqmCSterms}. Since this current does not carry any conserved quantum numbers, it need not satisfy the bound~\eqref{VW bound}, due to the ``disconnected diagrams'' reviewed above. 

In order to understand whether these can actually lead to spontaneous symmetry breaking for the~$U(1)_f$ symmetry, we should ask whether there can be a single-particle NGB pole at~$p^2 = 0$ in the momentum-space two-point function,
\begin{equation}
    \langle J_\mu^f(p) J_\nu^f(-p)\rangle~.
\end{equation}
The most natural possibility is that such a pole can arise from single-photon exchange. This requires the~$U(1)_m$ symmetry, with current
\begin{equation}
J^m_\mu = \frac{1}{2\pi} \varepsilon_{\mu\nu\rho} \partial^\nu a^\rho \,,  
\end{equation}
to also be spontaneously broken. We will argue in section~\ref{section:strongVafaWitten} that this is indeed the case. To simplify the discussion we also assume that the photon is the only massless particle in the IR as long as the fermion mass~$m \neq 0$ does not vanish (for further comments on this assumption, see section~\ref{sec:caveats}).

The weakly-coupled IR effective Lagrangian for the photon NGB is given by~\eqref{qfqmCSterms}, which we repeat here,
\begin{equation}\label{generalqfqmMaxwell}
 \mathscr{L} = -{1 \over 2 e^2_m} \left|da\right|^2 + {q_f \over 2 \pi} \Hat{A}_f \wedge da + {q_m \over 2 \pi} A_m \wedge da~ + \cdots~.
\end{equation}
Here the ellipsis denotes (irrelevant) higher-derivative self-interactions of the photon. Recall from the discussion around~\eqref{qfqmCSterms} that the quantized Chern-Simons levels indicate the~$U(1)_f$ and~$U(1)_m$ charges~$(q_f, q_m)$ of the monopole operator that condenses, and that anomaly-matching requires both charges to be odd. We can now use the low-energy Maxwell Lagrangian~\eqref{generalqfqmMaxwell} to compute the leading IR behavior of all current two-point functions,\footnote{~The photon propagator is~$\braket{a_\mu (p) a_\nu (-p)} = - \frac{ie_{m}^2}{p^2} \eta_{\mu\nu}$, up to gauge-dependent terms.} 
\begin{equation}
\braket{J_\mu^a (p) J_\nu^b (-p)} = -\frac{ie_{m}^2}{(2\pi)^2} \left( \frac{p_\mu p_\nu}{p^2} - \eta_{\mu\nu} \right) q_a q_b~, \qquad a, b = f, m~. 
\end{equation}
All three correlators contain the single-particle NGB pole from the photon, but since the two charge~$q_f, q_m$ are both non-zero, there is precisely one linear combination of the currents for which the pole cancels,
\begin{equation}\label{unbroken}
q_m J_\mu^f - q_f J^m_\mu~.
\end{equation}
The corresponding symmetry~$q_m U(1)_f - q_f U(1)_m$ thus remains unbroken. This is precisely the stabilizer group of the condensing monopole with charges~$q_f, q_m$. The simple reason is that there are two~$U(1)$ symmetries, but only one photon that can serve as NGB. Note that this general discussion applies to the large-$m$ Coulomb phase (with~$m > 0$), where ~$q_f = q_m = 1$ and~$U(1)_- = \frac{1}{2}( U(1)_m - U(1)_f)$ is unbroken, in agreement with~\eqref{unbrokenqfqm1}.

\subsubsection{A Bound on Electrically Charged Matter}
\label{section:strongVafaWitten}

We would like to argue for the existence of a massless photon, posited above, by showing that all electrically charged degrees of freedom (fundamental or composite) decouple exponentially rapidly at long distances, as long as the fermion mass in~\eqref{vwmass} is positive, $m > 0$. To see this, consider any composite operator~$\CO_q(x)$ of gauge charge~$q$ under our dynamical Spin$_c$ gauge field~$a$, e.g.~we could take~$\CO_q \sim \psi^q$. The two-point function of~$\CO_q$ is given by
\begin{equation}
\left\langle\cO_q^\dagger(y)\exp\left(iq\int_x^y a\right)\cO_q(x)\right\rangle~,
\label{strongvw}
\end{equation}
where we have included a suitable charge-$q$ Wilson line to ensure gauge invariance. We now use the Vafa-Witten bound~\eqref{vwpropbound} on the electron propagator in a fixed photon background~$a$ that holds for any non-zero fermion mass~$m$, together with the fact that the Wilson line is a pure phase in Euclidean signature, to obtain the following uniform bound,
\begin{equation}
\left|\Big\langle\cO_q^\dagger(y)\exp\left(iq\int_x^y a\right)\cO_q(x)\Big\rangle\right| \lesssim e^{-  m |q| |x-y|}~.
\label{strongvwii}
\end{equation}
Here it is crucial that there are no disconnected diagrams of the sort reviewed below~\eqref{VW bound}, because the operator~$\CO_q$ carries non-zero electric charge.  

Several comments are in order:
\begin{itemize}
\item[(i)] The argument is similar in spirit to the one that Vafa and Witten gave~\cite{Vafa:1983tf} to show that baryon number symmetry is not spontaneously broken in QCD, where the baryon current two-point function is also afflicted by disconnected diagrams. However, the two-point functions of operators that carry baryon number must decay exponentially because gluons do not carry baryon number, i.e.~there are no ``disconnected diagrams.'' 

\item[(ii)] The bound is completely blind to the details of the Wilson line (e.g.~its shape), because it is a pure phase. (We could also consider more complicated electric flux configurations consistent with Gauss' law.) Since the bound is unchanged, it seems plausible to attribute the exponential decay in~\eqref{strongvwii} to massive electric charges at~$x$ and~$y$, rather than the electric flux configuration extending between them.\footnote{~TD thanks Max Metlitski for a useful discussion about this point.}

\item[(iii)] The bound~\eqref{strongvwii} not only shows that the charged electrons decouple at long distances. It also rules out the existence of non-perturbative electrically charged massless bound states. This is reasonable, given the Coulomb repulsion between the charged constituents of such putative bound states. 

A special case is that there are no composite scalar Higgs fields (necessarily with even~$q$) that can condense and Higgs the~$U(1)$ gauge group to its~$\Z_q$ subgroup. Precisely such a phase is engineered in section~\ref{section: TQFT deformation}, and shown to match all anomalies, by introducing a fundamental Higgs field~$h$ of charge~$q$ and giving it a vev. Here we see that this scenario cannot arise dynamically in QED$_3$ as long as~$m \neq 0$.
\end{itemize}

Since all electrically charged degrees of freedom -- fundamental and composite -- decouple at long distances, it is reasonable to conclude that the low-energy theory contains a (nearly) free photon for all~$m > 0$. Thus the theory is in a Coulomb phase with spontaneously broken~$U(1)_m$ symmetry, and the photon is the corresponding NGB.  This can be rephrased in terms of the emergent, continuous electric 1-form symmetry~\cite{Gaiotto:2014kfa} that is present in this phase. This symmetry is explicitly broken by the charged fermions in the UV theory, but it emerges below the charged matter gap that is present as long as~$m \neq 0$. 

As a sanity check, we compute the photon propagator in QED$_3$ with an even number~$N_f$ of charge-1 electrons, in the presence of a mass deformation that gives a mass $+m$ to $N_f/2$ of the fermions and a mass $-m$ to the other $N_f/2$. The Euclidean path integral of this theory has positive measure, so that the decoupling of electric matter deduced above for the~$N_f = 2$ theory continues to hold. The 1PI resummed 1-loop photon propagator in the presence of the mass deformation takes the following form (in Euclidean signature with metric $\delta_{\mu\nu}$),
\begin{equation}
\braket{a_\mu(p) a_\nu(-p)} =  \frac{e^2}{p^2 + N_f e^2 f(p^2,m^2)} \delta_{\mu\nu}\,,
\label{largeNpropagator}
\end{equation}
up to gauge-dependent $p_\mu p_\nu$ terms that we drop. Note that this answer is reliable for any~$N_f$ as long as~$m$ is sufficiently large; it becomes exact in the large-$N_f$ limit, with fixed $\Lambda = N_f e^2$, even when~$m \lesssim \Lambda$. The function~$f(p^2, m^2)$ is given by 
\begin{equation}
f(p^2,m^2) = \frac{1}{8\pi} \left( 2|m| + \frac{p^2-4m^2}{|p|}\arcsin\left({\frac{|p|}{\sqrt{|p|^2+4m^2}}}\right) \right)
\simeq
\begin{cases}
    \dfrac{p^2}{12 \pi |m|} &\text{ if } |p| \ll |m| \,, \\ \vspace{1mm}
    \dfrac{|p|}{16} &\text{ if } |p| \gg |m| \,.
\end{cases}
\end{equation}
Note that~$|p| \equiv \sqrt{p^2} \geq 0$ in Euclidean signature. We thus see that for any finite $|m|$, the propagator in the deep IR has a single-particle NGB pole compatible with the general non-perturbative considerations above. Explicitly, as~$|p| \rightarrow 0$, we find that 
\begin{equation}
\braket{a_\mu(p) a_\nu(-p)} =  \frac{e^2_{m}}{p^2} \delta_{\mu\nu} + \CO(1) \,, \qquad \frac{1}{e^2_{m}} \equiv \frac{1}{e^2} + \frac{N_f}{12\pi|m|}~.
\label{ir limit propagator}
\end{equation}

\subsubsection{Caveates and Open Questions}\label{sec:caveats}

We have argued that any non-zero triplet mass~$\vec m \neq 0$ for the fermions leads to a nearly free photon NGB in the IR -- described by a low-energy action of the form~\eqref{generalqfqmMaxwell}, with quantized (in fact odd) Chern-Simons levels~$q_f, q_m$ describing the quantum numbers of the monopole operator that spontaneously breaks~$U(1)_f$ and~$U(1)_m$ to the subgroup discussed around~\eqref{unbroken}, in a way compatible with anomaly matching.  

Let us enunciate some of the gaps that remain between this statement and the stronger conjecture that the large-$|\vec m|$ Coulomb phase in section~\ref{subsection:largemphase}, smoothly persists for all~$\vec m \neq 0$:
\begin{itemize}
    \item The argument of section~\ref{section:strongVafaWitten} does not rule out the possibility of massless electrically neutral degrees of freedom in addition to the photon. Such degrees of freedom are absent at large~$|\vec m| \gg e^2$, and their appearance signals a second-order phase transition. Since they are electrically neutral, they should not renormalize the Chern-Simons levels~$q_f, q_m$ in~\eqref{qfqmCSterms}. 
    
    We can argue this as follows: the appearance of electrically neutral massless particles does not affect the electric 1-form symmetry of the photon that emerges in the IR (see above). This symmetry has (emergent) mixed 't Hooft anomalies with the~$U(1)_f$ and~$U(1)_m$ 0-form symmetries, with quantized anomaly coefficients~$q_f, q_m$. The latter are thus not affected by such a second-order phase transition.

    \item There can be first-order transitions across which the IR effective action and~$q_f, q_m$ jump abruptly, leading to a change in the pattern of spontaneous symmetry breaking. 
\end{itemize}

\subsection{Extrapolating to Symmetry Breaking at~$\vec m = 0$} \label{section:zeromass}

We will explore the implications of the phase diagram at non-vanishing fermion triplet mass~$\vec m \neq 0$ for the physics at the origin~$\vec m = 0$. For the purpose of this discussion, we will adopt the conjecture stated at the beginning of section~\ref{section: extending phase diagram}, namely that the Coulomb phase at large~$|\vec m| \gg e^2$ smoothly extends to all~$\vec m \neq 0$, and in particular to a small neighborhood of the origin. Then the residual symmetry that is present is spontaneously broken by the vev of the minimal~$q_m = 1$ monopole operator~$\CM^i$ of QED$_3$. This vev is aligned with~$\vec m$ via the Hopf map, as in~\eqref{zetaspinor}, which we repeat here
\begin{equation}\label{zetaspinorii}
    \cM^i = |\langle \CM\rangle| \zeta^i(\hat m) e^{i \sigma}~, \qquad \zeta^\dagger(\hat m) \vec \sigma \zeta(\hat m) = \hat m \equiv \vec m/ |\vec m|~.
\end{equation}
The IR theory only consists of a massless photon that furnishes the NGB for the spontaneously broken symmetry. 

Let us now contemplate the massless theory at the origin~$\vec m = 0$. The most minimal scenario -- which is manifestly consistent with the extrapolation of~\eqref{zetaspinorii} to~$\vec m = 0$ along all directions~$\vec m$ --  is the one proposed in section~\ref{introsquasheds3}: the monopole operator~$\CM^i$ acquires a vev~\eqref{mvevintro}, leading to the~$\t {S^3}$ sigma model with metric~\eqref{HopfMetricIntro}, and a vev~\eqref{OinIR} for the~$SU(2)_f$ triplet fermion bilinear~$\vec \CO = i \b \psi \, \vec\sigma \psi$ that is aligned with the monopole vev through the Hopf map. As we have already explained in section~\ref{section:thetaispiintro}, and will further elaborate in section~\ref{section: IR QED}, anomaly matching requires a~$\theta$-angle in the~$\t {S^3}$ sigma model, with coefficient~$\theta = \pi$. 

We will now argue that this symmetry-breaking scenario is the only plausible physical scenario, given what is already known about QED$_3$ with~$N_f = 2$ flavors. To this end, let us contemplate the possible alternatives:\footnote{~A very general loophole, which always afflicts extrapolations such as~$\vec m \to 0$, is that there may simply be unexpected/unnecessary degeneracies or vacua at~$\vec m = 0$ that are not protected by any symmetry. This is implausible in a strongly coupled theory, but it can happen if there is a suitably small/large parameter. An example of such accidental degeneracies in 3d large-$N$ QCD that are lifted at large but finite~$N$ was discussed in~\cite{Armoni:2019lgb, Argurio:2020her }.}
\begin{itemize}
\item[1.)] A scenario consistent with all constraints is that there is a gapless~CFT with unbroken~$U(2)$ global symmetry and unbroken time-reversal symmetry~$\CT$\footnote{~Note that the statements that~$\t \CT$ in~\eqref{Ttildedef} and~$U(2)$ are unbroken imply that~$\CT$ is also unbroken.} that must match the full anomaly in~\eqref{anomaly as c2}. As was already reviewed in section~\ref{secetion:qed3intro}, this scenario appears to be increasingly implausible in light of recent bootstrap constraints. We will therefore assume that it is not realized for~$N_f = 2$.\footnote{~Note that the scenario of a symmetry-preserving CFT is expected to be realized in QED$_3$ with~$N_f \geq 4$.} 

\item[2.)] If the~$U(2)$ symmetry is spontaneously broken, the only scenario that does not involve the condensation of any monopole operator, and thus unbroken~$U(1)_m$, is the spontaneous symmetry-breaking pattern
\begin{equation}
    U(2) \qquad \longrightarrow \qquad {U(1)_f \times U(1)_m \over \Z_2}~.
\end{equation}
This scenario was already discussed at the end of section~\ref{massesVWintro}: it is precisely the breaking pattern associated with condensation of the~$SU(2)_f$ triplet fermion bilinear~$\vec \CO = i \b \psi \, \vec \sigma \, \psi$ (which we take, without loss of generality, to lie along the~$\vec e_3$ direction). It therefore leads, at low energies, to two NGBs described by a sigma model into~$\C\P^1$. 

Since there is an unbroken time-reversal symmetry at every point on the~$\C\P^1$ (which coincides with~$\t \CT$ in~\eqref{CtTonMonopole} at the north and south poles),  the anomaly~\eqref{U(1)anomalyii} remains non-trivial. Indeed, as we showed around~\eqref{mdotnintro}, the~$\C\P^1$ model can be trivially gapped by a~$\left(U(1)_f \times U(1)_m\right)/\Z_2$-preserving mass term~\eqref{mzmass}. There must then be an additional dynamical sector with~unbroken~$\left(U(1)_f \times U(1)_m\right)/\Z_2$ and~$\t \CT$ symmetry that matches the anomaly at each point of the~$\C\P^1$. This sector would then be fibered over the~$\C\P^1$, much as the Hopf fiber~$\sigma$ of our~$\t S^3$ sigma model with metric~\eqref{HopfMetricIntro} is fibered over the~$\C\P^1$ base. 

There are two possibilities for this dynamical sector:
\begin{itemize}
\item[2a.)] It could be gapped, with a non-invertible TQFT at low energies that matches the anomaly via symmetry fractionalization (see~\cite{Brennan:2022tyl,Delmastro:2022pfo} for a recent discussion). An explicit example of this kind, with further details, is described in section~\ref{section: TQFT deformation}. Note, however, that this scenario is not consistent with the constraint that turning on a small triplet mass~$\vec m$ leads to a Coulomb phase with spontaneously broken~$U(1)_m$.

\item[2b.)] It could be a gapless, symmetry-preserving CFT with~$\left(U(1)_f \times U(1)_m\right)/\Z_2$ and~$\t \CT$ symmetries that matches the anomaly. In this scenario, the~$U(1)_f$-preserving fermion mass~$\vec m = m \vec e_3$ can flow to a non-trivial operator in the CFT that preserves all of its symmetries and drives it into a Coulomb phase. While this scenario is, strictly speaking, compatible with our general constraints, it involves an unnatural tuning: the massless CFT point has to emerge at exactly~$m = 0$; but from the point of view of the CFT this value of the mass is (in the absence of symmetry-enhancement) not in any way singled out. 

\end{itemize}

\item[3.)] The only remaining scenario is that~$\left(U(1)_f \times U(1)_m\right)/\Z_2$ is further broken to a subgroup by a monopole operator whose quantum numbers are fixed by considering small triplet-mass deformations~$\vec m \neq 0$. This leads to our proposed symmetry-breaking scenario~$\langle \CM^i\rangle \neq 0$. 

\end{itemize}

\section{Candidate Phases and Anomaly Matching in the IR}
\label{section: IR QED}

Here we elaborate on the discussion in section~\ref{section:deformationIntro}. In particular, we will give a complementary point of view on the~$\theta$-angle (with~$\theta = \pi$) that we argued in section~\ref{section:thetaispiintro} is needed to match the anomalies.

\subsection{Deforming QED$_3$ to the~${\t S^3}$ Sigma Model}
\label{section: Deformation argument}

Many aspects of our proposed symmetry-breaking phase -- in particular the intricacies of anomaly matching -- can be understood by explicitly realizing it in a deformed version of QED$_3$ that preserves all of its symmetries and anomalies and can be analyzed at weak coupling.\footnote{~This strategy was also employed in \cite{Cordova:2018acb,Dumitrescu:2023hbe} to explore subtle aspects of various gauge-theory phases.} 

\subsubsection{Adding a Scalar Field}

We introduce an elementary (gauge-neutral) real scalar field~$\vec \phi$ in the adjoint representation of~$SU(2)_f$
\begin{equation}
\vec \phi = (\phi^I)_{I = 1, 2, 3}~, \qquad (\phi^I)^\dagger = \phi^I~.
\end{equation}
In addition to canonical kinetic terms for~$\vec \phi$, which we add to the QED$_3$ Lagrangian \eqref{QEDLagrangian}, we further deform the theory by the following Yukawa couplings,\footnote{~In principle we could multiply our Yukawa couplings by an arbitrary coupling constant~$y > 0$, but it plays no significant role in our discussion so that we simply take~$y =1$.} 
\begin{equation}
\mathscr{L}_{\rm Y} = i \, \vec \phi \cdot \left(\Bar \psi \vec \sigma \psi\right) = i  \, \phi^I \Bar\psi_i (\sigma^I)^{i}_{~ j} \psi^j \,,
\label{Yukawa L}
\end{equation}
and a scalar potential for~$\vec \phi$,
\begin{equation}
V_{\phi} = -\mu^2 \phi^I \phi^I + \lambda (\phi^I \phi^I)^2 \,, \qquad \text{ with }\mu^2,\,\lambda > 0 \,.
\label{scalar pot}
\end{equation}
We give $\vec \phi$ the same symmetry action as the triplet mass, so that this theory has the full $G_\text{UV}$ symmetry as well as time reversal~$\cT$. Crucially, this means that all anomalies that are present in QED$_3$ must also be matched.

If we take the mass parameter~$\mu$ in the scalar potential~$V_\phi$ to be very large, $\mu \gg e^2$, then~$\vec \phi$ gets a large vev,
\begin{equation}\label{philargevev}
|\braket{\vec \phi}| = v \sim \mu / \sqrt \lambda \gg e^2~.
\end{equation}
This leads to spontaneous symmetry breaking of $SU(2)_f$ to $U(1)_f$, with two massless NGBs and a $\mathbb{CP}^1 = SU(2)_f/U(1)_f$ target space parametrized by
\begin{equation}
    \vec n = {\vec \phi \over v}, \qquad \vec n^2 = 1~.
\end{equation}
Due to the Yukawa couplings, the fermions get a non-degenerate mass at each point of the target space and can be integrated out.

Thus, the low-energy Lagrangian is superficially just given by the two-derivative kinetic terms for the~$\bC\bP^1$ nonlinear sigma model and the photon, 
\begin{equation}\label{IRkinLag}
    \mathscr{L}^\text{IR}_\text{kinetic} = -{v^2 \over 2} |\partial_\mu \vec n|^2 - {1\over 4 e^2} f^{\mu\nu} f_{\mu\nu} + \cdots~.
\end{equation}
Here the ellipsis denotes terms of higher than second order in the derivative expansion that we are not keeping track of. 

\subsubsection{An Important Chern-Simons Term}

We will now show that this Lagrangian is incomplete, because it is missing an important Chern-Simons term that couples the sigma model and the photon already at the two-derivative level. The full IR Lagrangian is instead given by adding to~$\mathscr{L}^\text{IR}_\text{kinetic}$ in~\eqref{IRkinLag} a Chern-Simons-like term, which is schematically
\begin{equation}\label{CSterm}
        \mathscr{L}^\text{IR}_\text{CS} = - a \wedge n^*\Omega_2 \,,
\end{equation}
where
\begin{equation}\label{omegadef}
     \Omega_2 = {1 \over 8 \pi} \varepsilon_{IJK} n^I dn^J \wedge dn^K~, \qquad \int_{\bC\bP^1} \Omega_2 = 1~,
\end{equation}
is the unit volume form on $\mathbb{CP}^1$. This is a two-derivative term, and crucially, its presence means the Skyrmion current $n^*\Omega_2$ is gauged.

This Chern-Simons term can be deduced by studying two Dirac fermions with the triplet mass $\vec \phi$ winding around $\mathbb{CP}^1$ at infinity, with $a$ treated as a background field. It was computed in various places, such as \cite{Abanov_2000,Hsin_2020,Choi_2022}. We will give an even simpler one-loop derivation below; we will also see that this term (suitably completed) is responsible for anomaly matching.

Before we do so, we will need to define~\eqref{CSterm} precisely, in a way that is also manifestly time-reversal invariant. We choose a 4-manifold $\cM_4$ whose boundary is spacetime, $\partial \cM_4 = \cM_3$, as well as an extension of the Spin$^c$ structure $a$ and the $\mathbb{CP}^1$ map to the bulk.\footnote{~Such an extension always exists, since the bordism group of Spin$^c$ 3-manifolds with a map to $\mathbb{CP}^1$, $\Omega_3^{{\rm Spin}^c}(\mathbb{CP}^1) = \Omega_3^{{\rm Spin}^c} \oplus \Omega_1^{{\rm Spin}^c} = 0$.} We define the term in the path integral to be
\begin{equation}\label{CP1CStermdef}
    \exp\left(-i \int_{\cM_4} da \wedge n^*\Omega_2\right).
\end{equation}
This is independent of the extension $\cM_4$ because on a closed $\cM_4$ (obtained by gluing two such extensions) this integral reduces, using the integrality of $n^*\Omega_2$, the fact that $\oint da = \pi w_2(T\cM_4)$, and the Wu formula, to
\begin{equation}
    \begin{split}
        \exp\left(-i \oint_{\cM_4} da \wedge n^*\Omega_2\right) = \exp\left(-i\pi \oint_{\cM_4} w_2(T\cM_4) \cup n^*\Omega_2\right) \\
        = \exp\left(-i\pi \oint_{\cM_4} n^*\Omega_2 \wedge n^*\Omega_2\right) = 1 \,,
    \end{split}
\end{equation}
where in the last line we used $n^*\Omega_2 \wedge n^* \Omega_2 = n^*(\Omega_2 \wedge \Omega_2) = 0$, since $\mathbb{CP}^1$ is 2-dimensional.

We will see below in Section \ref{section: duality with S3} that for $\cM^3 = S^3$, the quantity \eqref{CP1CStermdef} is $(-1)^\text{Hopf number}$. However, in general this term depends non-trivially on $a$. Another way to see it, which was described in \cite{Freed_2018}, is by choosing a $U(1)$ connection $\alpha$ on $\mathbb{CP}^1$ with $d\alpha = 2\pi \Omega_2$, and defining \eqref{CP1CStermdef} as the Chern-Simons term $- \frac{1}{2\pi} a \wedge dn^*\alpha + \frac{1}{4\pi} n^*\alpha \wedge dn^*\alpha$. This can be defined in a conventional way, by treating $n^*\alpha$ as a standard $U(1)$ connection and extending it as such to $\tilde \alpha$ on $\cM_4$. We then compute
\begin{equation}
\exp\left[ i\int_{\cM_4} \left( -\frac{1}{2\pi}  da \wedge d\tilde \alpha + \frac{1}{4\pi} d\tilde \alpha \wedge d\tilde \alpha  \right) \right] \,.   
\end{equation}
This equals \eqref{CP1CStermdef}, since we can choose the extension $n^*\alpha$ of $\alpha$ to $\cM_4$ corresponding to our extension of $n$, which makes the second term vanish. Our new expression \eqref{CP1CStermdef} makes time-reversal symmetry explicit.\footnote{~This term gives a generator of Anderson dual cobordism $\tilde \Omega^3_{{\rm Spin}^c}(S^2) = \Omega^1_{{\rm Spin}^c} = \bZ$, which we will also verify below.}

\subsubsection{Background Fields and Anomaly Matching}

Let us now show that this term is responsible for matching the anomaly in $\mathscr{L}_{\rm IR}$. To do this, we turn on the~$SU(2)_f$ background gauge fields~$A_f^I$ and the~$U(1)_m$ background gauge field~$A_m$ as in \eqref{QEDlagrangianbackground}. The appropriate~$SU(2)_f$ covariantization of~$n^*\Omega_2$ is given by (see section 3.3 of~\cite{Cordova:2018acb}) 
\begin{equation}\label{covOmega}
\tilde \Omega_2 = {1 \over 8 \pi}\left(\varepsilon_{IJK} n^I d_{A_f} n^J \wedge d_{A_f} n^K - 2 n^I F_f^I\right)~.
\end{equation}
This is closed and~$SU(2)_f$ invariant, but it can have fractional periods. As long as the background field~$A_f$ is a genuine~$SU(2)_f$ background, the periods of~$\tilde \Omega_2$ remain integral, because the space of~$SU(2)_f$ connections~$A_f$ modulo gauge transformations is connected. However this is not true if~$A_f$ is an~$SO(3)_f$ connection. One way to see this is to sit in a vacuum where~$n_I$ is a fixed constant vector, which we take to be aligned with the~$I = 3$ direction and to only activate the background gauge field~$A_f^{I = 3} = {1 \over 2}\hat A_f$ associated with the unbroken~$U(1)_f$ Cartan subgroup,
\begin{equation}\label{CartanOmega}
    n^I = \delta^{I3} \qquad \longrightarrow \qquad \tilde \Omega_2 = - {1 \over 4 \pi} F^{I = 3}_f = -{1 \over 2\pi} d \hat A_f~.
\end{equation}
Recalling section~\ref{section: Anomalies}, this shows that
\begin{equation}\label{nperiods}
\int_{\Sigma_2} \tilde \Omega_2 = {1 \over 2} \int_{\Sigma_2} w_2(SO(3)_f) = \int_{\Sigma_2} {d A_m \over 2 \pi} \,  \quad \text{mod} \, \bZ~.
\end{equation}
See also Appendix \ref{SO3equivappendix}.

We can thus write the full two-derivative IR effective Lagrangian coupled to background fields as follows (up to higher-derivative terms indicated by ellipses),
\begin{equation}\label{IRlagBF}
\mathscr{L}_\text{IR}[A_f, A_m] = 
-{v^2 \over 2} \left|d_{A_f} n^I\right|^2 - {1 \over 4 e^2} f^{\mu\nu} f_{\mu\nu} - a \wedge \tilde \Omega_2 + {1 \over 2\pi} da \wedge A_m + \cdots~.
\end{equation}
The only terms that are not manifestly invariant under (dynamical or background) gauge transformations are the Chern-Simons terms. To define them precisely, we choose a 4-manifold $\cM_4$ and an extension of $a$, the $\mathbb{CP}^1$ field $n$, and the $U(2)$ background field $(A_f,A_m)$ to it. The Chern-Simons terms contribute the following path integral weight:
\begin{equation}
    \exp \left(i \int_{\cM_4} da \wedge \left(\frac{dA_m}{2\pi} - \tilde \Omega_2 \right) \right),
\end{equation}
where, as above, $\tilde \Omega_2$ is the equivariantization of $n^*\Omega_2$ using the extended background fields. If we turn them off, this reduces to \eqref{CP1CStermdef}.

However, now this weight \emph{does} depend on $\cM_4$ and the choice of extension of background fields, which is precisely the anomaly. We can see this by taking $\cM_4$ to be closed, for which we find
\begin{equation}\label{c2identity}
\begin{split}
    \exp \left(i \oint_{\cM_4} da \wedge \left(\frac{dA_m}{2\pi} - \tilde \Omega_2 \right) \right) &= \exp \left(i \pi \oint_{\cM_4} \left(w_2(T\cM_4) + 2 \frac{dA_m}{2\pi}\right) \cup \left(\frac{dA_m}{2\pi} - \tilde \Omega_2 \right) \right)
    \\ &= \exp \left(i \pi \oint_{\cM_4} c_2(U(2)) \right).
\end{split}
\end{equation}
The last equality is shown in Appendix \ref{appendixU2equivS2coh}.

\subsubsection{A One-Loop Derivation of the Chern-Simons Term}

The preceding discussion also offers a path to derive from first principles the presence of the Chern-Simons term~\eqref{CSterm}, including its precise coefficient. To this end, let us expand the IR Lagrangian~\eqref{IRlagBF} around the vacuum~$n^I = \delta^{I3}$ at the north pole of the~$\bC\bP^1$, and restrict to the background fields~$A_m, \hat A_f$. Using~\eqref{CartanOmega}, we get
\begin{equation}\label{amafIRlag}
\mathscr{L}_\text{IR}[A_f, A_m] = 
-{v^2 \over 2} \left|d_{A_f} n^I\right|^2 - {1 \over 4 e^2} f^{\mu\nu} f_{\mu\nu} + {1 \over 2\pi} a \wedge d (A_m + \hat A_f)~, \qquad v > 0~.
\end{equation}
Note that the monopole of the low-energy Maxwell theory is precisely charged under~$A_+ = A_m + \hat A_f$, while it is neutral under~$A_-$. Consequently, the monopole of the UV theory that condenses in this vacuum is precisely~$\cM^{i = 1}$, i.e.
\begin{equation}
    \langle \cM^i\rangle \sim \delta^{i1}~.
\end{equation}
This shows that the symmetry-breaking pattern is actually the one indicated in~\eqref{SSBintro}. 

In order to compute the mixed Chern-Simons term involving~$\hat A_f$ and~$a$ in~\eqref{amafIRlag}, we simply note that substituting~$\phi^I = v\, \delta^{I3}$ into the Yukawa couplings~\eqref{Yukawa L} leads to a triplet fermion mass~$m_3 = v > 0$. Since the fermions~$\psi^1$ and~$\psi^2$ both have gauge charge 1, but~$U(1)_f$ charges~$+1$ and~$-1$, respectively, it follows that the induced Chern-Simons term is exactly given by the usual 1-loop formula, by which each fermion contributes~$+1/2$ to the Chern-Simons level, leading to the level 1 Chern-Simons term in~\eqref{amafIRlag}. This is, in essence, exactly the same one-loop computation that we did in the large-$|\vec m|$ analysis of QED$_3$ in section~\ref{subsection:largemphase}, except that the mass~$\vec m$ is replaced by the dynamical scalar field~$\vec \phi$, whose angular part~$\vec n$ provides the massless~$\C\P^1$ degrees of freedom.

Note that if we allow the sign of~$v$ to be negative, then the Chern-Simons term for~$\hat A_f$ would flip sign, leading to
\begin{equation}\label{IRlagvnegative}
\mathscr{L}_\text{IR}[A_f, A_m] = 
-{v^2 \over 2} \left|d_{A_f} n^I\right|^2 - {1 \over 4 e^2} f^{\mu\nu} f_{\mu\nu} + {1 \over 2\pi} a \wedge d (A_m - \hat A_f)~, \qquad v < 0~.
\end{equation}
This shows that the monopole that condenses is charged under~$A_-$ and neutral under~$A_+$, which means that it must be~$\cM^{i = 2}$. The fact that the monopole vevs, and the unbroken~$U(1)$ subgroup is fibered over the~$\bC\bP^1$ base in this fashion is an inescapable consequence of the symmetry-breaking pattern~\eqref{SSBintro} triggered by the condensation of the fundamental monopole~$\cM^i$. 

\subsection{A Candidate $\C\P^1$ + TQFT Phase with Unbroken~$U(1)_m$}
\label{section: TQFT deformation}

It is instructive to contemplate other phases that have the same symmetries and anomalies as our model. These are in principle candidate phases for~QED$_3$, but are not in fact realized due to our arguments in section \ref{section:zeromass}. However they could conceivably play a role once QED is further deformed, as we are doing here.

Let us contemplate adding an extra scalar Higgs field~$h$ of electric charge~$2$ that is invariant under all global symmetries. To see that this is consistent with all our selection rules, note that~$h$ has the same quantum numbers as
\begin{equation}\label{Hqn}
h \sim   \varepsilon^{\alpha\beta} \psi_\alpha \vec \sigma \psi_\beta \cdot \vec\phi~.
\end{equation}
Then we can simply condense~$h$ by adding a suitable Higgs potential for it.\footnote{~Note that~$h$ in principle has its own~$U(1)$ global symmetry that only rotates it and nothing else, but we will not track it. To justify this we can imagine explicitly breaking this symmetry by adding a sufficiently small perturbation (e.g.~an irrelevant operator with a suppressed coefficient) that preserves all the symmetries of QED$_3$.} Then the~$U(1)$ gauge symmetry of~$a$ is Higgsed down to its~$\bZ_2$ subgroup, leading to a gapped phase with unbroken~$U(1)_m$ symmetry and a (non-invertible) topological~$\Z_2$ gauge theory that matches the anomaly. Note indeed that a~$\bZ_2$ TQFT has two different~$\Z_2^{(1)}$ 1-form symmetries, with a mixed 't Hooft anomaly
\begin{equation}
S_\text{bulk}[B_\pm] = \pi \int_{\cM_4} B_+ \cup B_-~, \qquad B_\pm \in H^2(\cM_4, \bZ_2)~,
\end{equation}
which matches the anomaly in~\eqref{anomalyAmAf} if we take~$B_\pm = d A_\pm/2\pi$.\footnote{~This means that the anomaly is matched by symmetry fractionalization in the TQFT phase, see for instance \cite{Brennan:2022tyl,Delmastro:2022pfo}.}

This shows that, at the level of anomaly matching, an acceptable scenario is the symmetry-breaking pattern~$U(2) \to (U(1)_f \times U(1)_m) / \Z_2$, with a~$\bZ_2$ TQFT fibered over the~$\bC\bP^1$ sigma model, to match the anomaly in the residual unbroken~$(U(1)_f \times U(1)_m) / \Z_2$ symmetry.

It is straightforward to generalize the discussion above to a Higgs field~$h$ of any even charge~$q$, leading to a~$\Z_q$ gauge theory that matches the anomaly.

\subsection{Recovering the~$\t {S^3}$ Sigma Model with~$\theta = \pi$ from Duality}
\label{section: duality with S3}

We will now return to the deformation analysis initiated in section~\ref{section: Deformation argument} and re-derive from that point of view the squashed~$\t {S^3}$ sigma model introduced in section~\ref{introsquasheds3} -- importantly including the~$\theta = \pi$ term in~\eqref{s3thetaintro} that is needed for anomaly matching. 

To this end, let us revisit the full IR Lagrangian~\eqref{IRlagBF} that describes the coupling of the~$\bC\bP^1$ model to Maxwell theory via a Chern-Simons term, and we perform a version of the standard Abelian duality transformation of Maxwell theory into the dual photon~$\sigma$ that shifts under~$U(1)_m$.

Explicitly, we write the theory in \eqref{IRlagBF} as follows (here it will be convenient to switch to Euclidean signature),
\begin{equation}
\mathscr{L}_{\rm IR}  =  \frac{1}{2e^2} |db - dA_0|^2 + \frac{i}{2\pi} db \wedge n^* \alpha - \frac{i}{2\pi} dA_0 \wedge n^* \alpha \,,
\end{equation}
where $A_0$ is a reference Spin$^c$ structure, $b$ is an ordinary $U(1)$ gauge field, related to our dynamical Spin$^c$ structure as $a = b - A_0$. The theory should not depend on the choice of $A_0$, and should also be invariant under gauge transformations of $\alpha$ and $b$. This is guaranteed by treating the Chern-Simons terms according to the prescription in section \ref{section: Deformation argument}. Here, this corresponds to treating $\frac{i}{2\pi} db \wedge n^* \alpha$ as a usual mixed Chern-Simons term of $U(1)$ gauge fields, but treating $- \frac{i}{2\pi} dA_0 \wedge n^* \alpha$ in a special way, extending both $A_0$ and $n$ to a 4d bulk $\cM_4$, and computing the path integral weight as
\begin{equation}\label{S2spincbordismterm}
    \exp\left(-i \int_{\cM_4} dA_0 \wedge n^*\Omega_2\right).
\end{equation}

We now proceed to dualize~$b$ as in ordinary 3d Maxwell theory. We let $db$ be an arbitrary 2-form $\lambda$, and introduce the dual $2\pi$-periodic field $\sigma$ as a Lagrange multiplier which sets $d\lambda=0$ and quantizes its periods:
\begin{equation}
\mathscr{L}_{\rm IR}  =    \frac{1}{2e^2} |\lambda - dA_0|^2 + \frac{i}{2\pi} \lambda \wedge (n^* \alpha - d\sigma) - \frac{i}{2\pi} dA_0 \wedge n^* \alpha \,.
\end{equation}
We have combined the $\sigma$ term with the ordinary Chern-Simons term, since now we cannot use the 4d prescription to define it. It is only gauge invariant by making $\sigma$ transform as with charge 1 under gauge transformations of $\alpha$, so that $(n^* \alpha - d\sigma)$ is a gauge-invariant, globally well-defined form.
Integrating out $\lambda$ yields the dual theory:
\begin{equation}\label{S3duallagrangian}
\Tilde{\mathscr{L}}_{\rm IR} = \frac{e^2}{8\pi^2} |d\sigma - n^*\alpha|^2 - \frac{i}{2\pi} dA_0 \wedge (d\sigma - n^*\alpha) - \frac{i}{2\pi} dA_0 \wedge n^* \alpha \,.
\end{equation}
Because $\sigma$ carries charge 1 under $n^*\alpha$ gauge transformations, it is a section of the $S^1$ bundle on $\cM_3$ obtained by pulling back the Hopf bundle $\t {S^3} \to \mathbb{CP}^1$ by $n$. Together, $n$ and $\sigma$ thus combine into an $\t {S^3}$ field $f(n,\sigma)$, where $e^2$ is the squashing parameter of this $\t {S^3}$ sigma model.

We will now show that the two topological terms in \eqref{S3duallagrangian} together are equal to a $\theta = \pi$ term in the path integral,
\begin{equation}\label{S3thetapi}
    \exp\left(i \pi\int_{\cM_3} f^*\Omega_3\right) \,,
\end{equation}
where $\Omega_3$ is the unit volume form on $\t {S^3}$. First, we show that they are independent of the choice of $A_0$. If we made a different choice $A_1$, the difference $A' = A_0 - A_1$ would be a $U(1)$ connection, and the change in the Lagrangian would be
\begin{equation}
    - \frac{i}{2\pi} dA' \wedge (d\sigma-n^*\alpha) - \frac{i}{2\pi} dA' \wedge n^*\alpha \,.
\end{equation}
In this expression, we can treat $\frac{i}{2\pi} dA' \wedge n^*\alpha$ as an ordinary mixed Chern-Simons term, and as such the two cancel. The $\sigma$ term meanwhile gives a $2\pi i$ integer because $\frac{dA'}{2\pi}$ has integer periods and thus contributes trivially.

This allows us to choose $A_0$ in fact to be a spin structure (since all orientable 3-manifolds admit one). This has $dA_0 = 0$, in which case the first topological term in \eqref{S3duallagrangian} vanishes, and the total path integral weight from the topological terms is given by \eqref{S2spincbordismterm}, where we extend $A_0$ as a Spin$^c$ connection. An advantage of this prescription is that \eqref{S2spincbordismterm} is a bordism invariant for spin 3-manifolds equipped with a map to $\mathbb{CP}^1$. This bordism group is $\Omega_3^{\rm Spin}(\mathbb{CP}^1) = \bZ_2$, generated by the $\t {S^3}$ equipped with the Hopf map and any spin structure.

The $\theta = \pi$ term \eqref{S3thetapi} is also a bordism invariant, but of spin (or even just unoriented) 3-manifolds with a map to $\t {S^3}$. This bordism group is $\Omega_3^{\rm Spin}(S^3) = \Omega_3^{SO}(S^3) = \bZ$, generated by $S^3$ with the identity map to itself and any spin structure. The map which sends a spin manifold $\cM_3$ equipped with a map $f:\cM_3 \to S^3$ to the same spin manifold equipped with the map $f \circ h:\cM_3 \to S^2$, where $h$ is the Hopf map, is thus reduction mod 2:
\begin{equation}
    \Omega_3^{\rm Spin}(S^3) = \bZ \qquad \to \qquad \Omega^{\rm Spin}_3(S^2) = \bZ_2 \,.
\end{equation}
The $\theta$-term $\exp({i \theta \int_{\cM_3} \Omega_3})$ parametrizes ${\rm Hom}(\Omega_3^{\rm Spin}(S^3),U(1)) = U(1)_\theta$ with $2\pi$ periodicity. It follows right away that the generator of ${\rm Hom}(\Omega^{\rm Spin}_3(S^2),U(1)) = \bZ_2$ corresponds to $\theta = \pi$.

We just need to check therefore that \eqref{S2spincbordismterm} is non-trivial on a generator of $\Omega_3^{\rm Spin}(S^2)$. We can choose as generator $\cM_3 = S^3$ equipped with the Hopf fibration $n:S^3 \to \mathbb{CP}^1$, and its unique spin structure. Recall now that $\mathbb{CP}^2$ is obtained by attaching a 4-ball $B^4$ with its 3-sphere boundary to a 2-sphere $S^2$ via the Hopf fibration (see e.g.~\cite{hatcher2005algebraic} page 7). If we remove another small 4-ball $B^4_\epsilon$ from the center of the $B^4$, we obtain a manifold $\cM_4$ with boundary $S^3$. We choose on this manifold a Spin$^c$ structure (note that $\mathbb{CP}^2$ is not Spin) $A_0$, which has all of its curvature supported in a compact neighborhood around the $S^2 \subset \cM_4$. The map $\cM_4 \to S^2$ comes from extending the Hopf map to $B^4 - B^4_\epsilon = S^3 \times I$ by the identity on $I$, and gluing this map to the identity map on $S^2$. Because of this, $\int_{S^2} n^*\Omega_2 = 1$. Although $n$ does not extend to all of $\mathbb{CP}^2$, the form $n^* \Omega_2$ does extend to a closed form $\beta_2$, also with $\int_{S^2} \beta_2 = 1$, thus representing the generator of $H^2(\mathbb{CP}^2,\bZ)$. Since all the curvature of $A_0$ is concentrated along this $S^2$, we can close the integral and obtain
\begin{equation}
    \exp\left(-i \int_{\cM_4} dA_0 \wedge n^*\Omega_2\right) = \exp\left(-i \pi \int_{\mathbb{CP}^2} \beta_2 \wedge \beta_2 \right) = -1 \,.
\end{equation}

We can do a sanity check by computing the anomaly in the dual $\t {S^3}$ theory. There is a very easy way to do this with equivariant cohomology, which says that there is an extension of the theta-term $\theta \, \Omega_3$ to a form $\theta \, \Tilde \Omega_3$ which satisfies $d \Tilde \Omega_3 = c_2(U(2))$ (see Appendix \ref{appendixS3withU2action}), thus realizing the theory at any $\theta$ in a gauge-invariant way with a $\theta \, c_2(U(2))$ term in a 4d bulk. For $\theta = \pi$, this agrees with what we computed above.

\section{Comments on $N_f>2$}
\label{section: Nf>2}

In this section, we extend our discussion to QED$_3$ with an even number $N_f=2n_f$ of Dirac fermions.
As we reviewed in the introduction, this theory is believed to flow to a non-trivial interacting CFT, which has been analyzed with a variety of approaches, including the numerical bootstrap, the large $N_f$ limit, and the $\varepsilon$ expansion. However, it is instructive to generalize our discussion to higher values of $N_f$, to determine which pattern of spontaneous symmetry breaking is consistent with the strong Abelian non-renormalization constraints explored in section \ref{section:strongVafaWitten}, anomaly matching, and the general idea that symmetry breaking in QED$_3$ is driven by the condensation of monopole operators.

In the scenario where symmetry breaking is driven only by fermion bilinears, the $PSU(2n_f) = SU(2n_f) / \bZ_{2n_f}$ flavor part of the global symmetry is spontaneously broken according to the pattern
\begin{equation}\label{SSBgenericNf}
    PSU(2n_f) \quad \longrightarrow \quad \frac{SU(n_f) \times SU(n_f) \times U(1)_f}{\bZ_{n_f} \times \bZ_{n_f} \times \bZ_{2}} \,,
\end{equation}
by the condensation of a fermion bilinear
\begin{equation}
\cO =\sum_{i=1}^{n_f} \left( \Bar\psi_i \psi^i - \Bar\psi_{i+n_f} \psi^{i+n_f} \right) \,,      
\end{equation}
and the $U(1)_m$ magnetic symmetry is unbroken.
In \eqref{SSBgenericNf}, the $U(1)_f$ subgroup acts with charge $+1$ on $\psi^i$ and $-1$ on $\psi^{i+n_f}$ ($i=1,\dots,n_f$).
The quotient is by the gauge transformations $(e^{2\pi i/n_f}\bI_{n_f},e^{2\pi i/n_f}\bI_{n_f},1)$ (generating a $\bZ_{n_f}$) and $(\bI_{n_f},\bI_{n_f},-1)$ (generating the $\bZ_2$) and by a $\bZ_{n_f}$ $(e^{2\pi i/n_f}\bI_{n_f},e^{-2\pi i/n_f}\bI_{n_f},e^{2\pi i/n_f})$. For $n_f = 1$ this reduces to $SO(3)_f = PSU(2)$ broken to $U(1)$, giving the $\mathbb{CP}^1$.
The fermion bilinear above also preserves a time reversal symmetry $\tilde \cT$ which is the naive $\cT$ in \eqref{T on fermions} times a $\pi/2$ flavor rotation $U_f$ rotating the first $n_f$ fermions into the second, analogous to \eqref{tildeT}. This $\t \cT$ does not commute with the broken flavor symmetries and so which time reversal is preserved depends on the vacuum we consider.

As for the $N_f=2$ case, this scenario is not compatible with the non-perturbative bound on electrically charged matter explored in section \ref{section:strongVafaWitten}. Upon deforming the theory with a time-reversal invariant mass, which we can choose to be $\mathscr{L}_m = im\cO$, we argued that this theory flows to a Coulomb phase with a massless photon for any $m\neq 0$. Thus, there is a condensing monopole operator which further breaks the magnetic $U(1)_m$ symmetry and the $U(1)_f$ symmetry in \eqref{SSBgenericNf} to a diagonal combination.

Let us briefly review the quantum numbers of monopole operators, following \cite{Borokhov:2002ib}, and the global symmetry for generic $N_f=2n_f$. The Hilbert space of zero modes of the free Dirac Hamiltonian on $\mathbb{R}_t \times S^2$, in a constant background of one unit of magnetic flux, is a Fock space with $2^{2n_f}$ degenerate spin-zero states. Imposing the Gauss law to select the gauge-neutral physical states, one gets that those are created by acting with exactly $n_f$ zero modes on the Fock vacuum, whose number is
\begin{equation}
\binom{2n_f}{n_f} \,. 
\end{equation}
They correspond to monopole operators $\cM^{i_1\dots i_{n_f}}$, transforming in the rank-$n_f$ antisymmetric representation of $SU(2n_f)$ and with unit $U(1)_m$ charge.
Note that a transformation by the $\bZ_{2n_f}$ center of $SU(2n_f)$ acts as
\begin{equation}
\cM^{i_1\dots i_{n_f}} \rightarrow (e^\frac{2\pi i}{2n_f})^{n_f} \cM^{i_1\dots i_{n_f}} = - \cM^{i_1\dots i_{n_f}} \,,
\end{equation}
which is a $\pi$ rotation of $U(1)_m$. Thus, the global structure of the (internal) symmetry is \cite{Benini:2017dus}
\begin{equation}
G_{UV} = \frac{SU(2n_f) \times U(1)_m}{\bZ_{2n_f}} \rtimes \bZ_2^\cC = \frac{U(2n_f)}{\bZ_{n_f}} \rtimes \bZ_2^\cC \,,
\label{Nf global sym}
\end{equation}
where the $\bZ_{2n_f}$ quotient is generated by $(e^{2\pi i/2n_f} \mathbb{I}_{2n_f},-1) \in SU(2n_f) \times U(1)_m$, which leads to the $\bZ_{n_f}$ quotient of $U(2n_f)$, generated by $e^{2\pi i/n_f} \mathbb{I}_{2n_f} \in U(2n_f)$.\footnote{~Indeed, $U(2n_f)$ is defined by taking the $\bZ_{2n_f}$ quotient generated by $(e^{2\pi i/2n_f} \mathbb{I}_{2n_f},e^{2\pi i/2n_f})$. Here, we are taking the $\bZ_{2n_f}$ quotient to be generated by $(e^{2\pi i/2n_f} \mathbb{I}_{2n_f},e^{i\pi})$, which is a further $\bZ_{n_f}$ identification.}
Under discrete symmetries, monopoles transform as \cite{Cordova:2017kue}
\begin{align}
\cC &: \cM^{i_1\dots i_{n_f}} \rightarrow (\cM^{i_1\dots i_{n_f}})^* \,, \\
\cT &: \cM^{i_1\dots i_{n_f}}(t) \rightarrow \frac{(-1)^{\frac{n_f(n_f-1)}{2}}}{n_f !}\varepsilon_{i_1 \dots i_{n_f}j_1\dots j_{n_f}}(\cM^{j_1\dots j_{n_f}})^*(-t) \,, \\
\cC\cT &: \cM^{i_1\dots i_{n_f}}(t) \rightarrow \frac{(-1)^{\frac{n_f(n_f-1)}{2}}}{n_f !}\varepsilon_{i_1 \dots i_{n_f}j_1\dots j_{n_f}}\cM^{j_1\dots j_{n_f}}(-t) \,,
\end{align}
so that in sectors with a non-trivial monopole number we have
\begin{equation}
\cC^2=1 \,, \qquad (\cC\cT)^2=\cT^2=
\begin{cases}
1 &\text{ if } N_f = 0 \text{ mod }4 \,, \\
(-1)^\cM  &\text{ if } N_f = 2 \text{ mod }4 \,.
\end{cases}
\end{equation}

Using a deformation argument analogous to the one considered in section \ref{section: Deformation argument}, it follows that monopole operators get a vev which aligns to the non-Abelian part of the symmetry-breaking pattern \eqref{SSBgenericNf}.\footnote{~Interestingly, this is also consistent with the analysis of the global minima of an $SU(2N)$-invariant quartic potential for a scalar field transforming in the rank-$N$ antisymmetric representation of $SU(2N)$: the negative-mass phase corresponds to the symmetry-breaking pattern $SU(2N) \rightarrow SU(N) \times SU(N)$ \cite{Kim:1980ec,Jetzer:1983ij}.}
Note that this is also dictated by the Vafa-Witten theorem, as the arguments of section \ref{section: Vafa-Witten} only allow for a non-trivial mixing between $U(1)_f$ and $U(1)_m$, which cannot be contaminated by the non-Abelian part of the unbroken global symmetry. Indeed, disconnected diagrams do not contribute to correlators of non-Abelian flavor currents, as they carry a non-trivial flavor charge.

Clearly, when the monopole operator condenses, it breaks the $U(1)_f$ and $U(1)_m$ global symmetries to a mixed $U(1)$. This will be fibered over space of bilinears in \eqref{SSBgenericNf}. For example, on the points pinned by the deformation $\mathscr{L}_m=im\cO$, we get that the monopole which condenses is
\begin{equation}
\cM^{1,\dots,n_f} \quad \text{ if } m>0 \,, \qquad \text{or} \qquad
\cM^{n_f+1,\dots,2n_f} \quad \text{ if } m<0 \,,
\label{Nf monopole vev}
\end{equation}
which is an immediate generalization of the scenario we proposed for $N_f=2$.
We thus have the symmetry-breaking pattern
\begin{equation}
 \frac{U(2n_f)}{\bZ_{n_f}} \rightarrow  \frac{SU(n_f) \times SU(n_f) \times U(1)}{\bZ_{n_f}\times\bZ_{n_f}} = \frac{SU(n_f) \times U(n_f)}{\bZ_{n_f}} \,,
 \label{genNfsymbreak}
\end{equation}
where in the first step on the right-hand side one $\bZ_{n_f}$ is generated by $(\bI_{n_f},e^{-2\pi i/n_f} \bI_{n_f}, e^{2\pi i/n_f}) \in SU(n_f) \times SU(n_f) \times U(1)$,\footnote{~The choice of which $SU(n_f)$ element is the identity depends on the vev (for definiteness here we considered the first case in \eqref{Nf monopole vev}), as we know from the fact that the broken $U(1)$ combination -- which is the one under which the monopole of the low-energy Maxwell theory has unit charge -- is also fibered over the base.} and allows us to rewrite the combination $(SU(n_f) \times U(1))/\bZ_{n_f}$ as $U(n_f)$, whereas the other $\bZ_{n_f}$ acts as the quotient on the left-hand side. The theory then flows to a sigma model with target space
\begin{equation}
\frac{U(2n_f)}{SU(n_f) \times U(n_f)} \,. 
\label{Nf ir sigmamodel}
\end{equation}
As a consistency check, for $n_f=1$ we get the symmetry-breaking pattern in \eqref{SSBintro} and an $S^3$ sigma model. 

As in the $N_f=2$ case, an analogous deformation argument automatically proves that this symmetry-breaking scenario is compatible with anomaly matching, and could in principle occur in some phase. It would be interesting to further investigate how the UV 't Hooft anomaly is matched by the IR sigma model with target space \eqref{Nf ir sigmamodel}. This must happen both for the pure anomaly of the global symmetry, which is always present for $N_f \geq 4$ due to the non-trivial $\bZ_{N_f/2}$ quotient in \eqref{Nf global sym}, and for the anomalies involving time-reversal, analyzed in \cite{Benini:2017dus,Hsin:2024abm}.

However, let us emphasize again that the massless theory is not expected to break the global symmetry as in \eqref{genNfsymbreak}, but it rather flows at low energies to a non-trivial strongly coupled CFT which preserves the whole global symmetry. As a sanity check, notice that the large-$N_f$ exact photon propagator \eqref{largeNpropagator} at $m=0$ behaves as $1/|p|$, signalling that the magnetic symmetry is indeed unbroken in the CFT.

\section*{Acknowledgments}
TD and PN are grateful to Ken Intriligator and Orr Sela for related discussions and collaboration. TD thanks Juan Maldacena for related collaboration, as well as Zohar Komargodski and Max Metlitski for useful discussions. PN thanks Lorenzo Di Pietro for many discussions about QED$_3$ and related collaborations. RT thanks Arun Debray, Cameron Krulewski, Yu Leon Liu, and Matt Yu for many enlightening discussions of cobordism. The authors acknowledge support by the Mani L. Bhaumik Institute for Theoretical Physics and by the Simons Collaboration on Global Categorical Symmetries. TD and PN are also supported by a DOE Early Career Award under DE-SC0020421. RT is supported by the Mani L. Bhaumik Presidential Term Chair in Theoretical Physics at UCLA.

\appendix

\section{Cobordism Calculations}
\label{appcobcalcs}

\subsection{$Pin^-(2)$}

First we need some of the cohomology of $Pin^-(2)$, which sits in the unique non-split, twisted extension
\[U(1) \to Pin^-(2) \to \Z_2.\]
The Lyndon–Hochschild–Serre spectral sequence (LHSSS) has $E_2^{p,q} = H^p(B\bZ_2,H^q(BU(1),\Z))$ where we can choose whether $\Z_2$ acts on the coefficients $\Z$. If the chosen twist is $\tau$ and the nontrivial one is $\sigma$, we get
\[E_2^{p,4n} = H^p(B\bZ_2,\Z^\tau) \\ E_2^{p,4n+2} = H^p(B\bZ_2,\Z^{\tau \otimes \sigma}).\]
Let us write $A$ for the generator of $H^1(B
\Z_2,\Z_2) = \Z_2$ and $c_1$ for the generator of $H^2(BU(1),\Z) = \Z$. For $\tau$ trivial we get
\begin{center}
\begin{tikzcd}[row sep=small, column sep = {3em,between origins}]
& 0 \\
& 0 & 0 \\
& \Z^{c_1^2} & 0 & \Z_2^{A^2 c_1^2} \\
& 0 & 0 & 0 & 0 \\
& 0 & \Z_2^{Ac_1} & 0 & \Z_2^{A^3 c_1} & 0 & \Z_2^{A^5 c_1} \\
& 0  & 0 & 0 & 0 & 0 & 0 & 0\\
q=0& \Z & 0 & \Z_2^{A^2} & 0 & \Z_2^{A^4} & 0 & \Z_2^{A^6} & 0 & \Z_2^{A^8} \\
& p=0
	\arrow[from=5-3, to=7-6]
 \arrow[from=5-5, to=7-8]
 \arrow[from=5-7, to=7-10]
\end{tikzcd}
\end{center}
The arrows indicate the differential $d_3(c_1) = A^3$, which comes from the non-trivial extension. This yields
\[\begin{matrix}
    H^0(BPin^-(2),\Z)= & \Z \\ H^1(BPin^-(2),\Z)= & 0 \\ H^2(BPin^-(2),\Z)= & \Z_2^{A^2} \\ H^3(BPin^-(2),\Z)= & 0 \\ H^4(BPin^-(2),\Z)= & \Z^{c_1^2} \\ H^5(BPin^-(2),\Z)= & 0 \\ H^6(BPin^-(2),\Z)= & \Z_2^{A^2 c_1^2}
\end{matrix}\]
For $\tau = \sigma$ we get
\begin{center}
\begin{tikzcd}[row sep=small, column sep = {3em,between origins}]
& 0 \\
& 0 & 0 \\
& 0 & \Z_2^{A c_1^2} & 0 \\
& 0 & 0 & 0 & 0 \\
& \Z^{c_1} & 0 & \Z_2^{A^2 c_1} & 0 & \Z_2^{A^4 c_1} & 0\\
& 0  & 0 & 0 & 0 & 0 & 0 & 0\\
q=0 & 0 & \Z_2^A & 0 & \Z_2^{A^3} & 0 & \Z_2^{A^5} & 0 & \Z_2^{A^7} \\
& p=0
	\arrow[from=5-2, to=7-5]
 \arrow[from=5-4, to=7-7]
 \arrow[from=5-6, to=7-9]
\end{tikzcd}
\end{center}
This yields
\[\begin{matrix}
    H^0(BPin^-(2),\Z^\sigma)= & 0 \\ H^1(BPin^-(2),\Z^\sigma)= & \Z_2^A \\ H^2(BPin^-(2),\Z^\sigma)= & \Z^{2 c_1} \\ H^3(BPin^-(2),\Z^\sigma)= & 0 \\ H^4(BPin^-(2),\Z^\sigma)= & 0 \\ H^5(BPin^-(2),\Z^\sigma)= & \Z_2^{A c_1^2} \\ H^6(BPin^-(2),\Z^\sigma)= &0
\end{matrix}\]

The $\tau$-twisted oriented cobordism groups of $Pin^-(2)$ we need can now be computed by the Atiyah-Hirzebruch spectral sequence (AHSS), which has $E_2^{p,q} = H^p(BPin^-(2),\Omega^q_{SO} \otimes \Z^\tau)$, where
\[\begin{matrix}
    \Omega^{-1}_{SO}= & \Z \\ \Omega^{0}_{SO}= & 0 \\ \Omega^{1}_{SO}= & 0 \\ \Omega^{2}_{SO}= & 0\\ \Omega^{3}_{SO}= & \Z \\ \Omega^{4}_{SO}= & 0 \\ \Omega^{5}_{SO}= & \Z_2.
\end{matrix}\]
For $\tau$ trivial we get
\begin{center}
\begin{tikzcd}[row sep=small, column sep = {3em,between origins}]
& 0  \\
& \Z^{p_1/3} & 0 \\
& 0 & 0 & 0  \\
& 0  & 0 & 0 & 0 \\
& 0  & 0 & 0 & 0 & 0 \\
q = -1 & \Z & 0 & \Z_2^{A^2} & 0 & \Z^{c_1^2} & 0 \\
& p=0
\end{tikzcd}
\end{center}
(with no possible differentials.) This gives
\[\begin{matrix}
    \Omega^{-1}_{SO}(BPin^-(2)) = & \Z \\ \Omega^{0}_{SO}(BPin^-(2))= & 0 \\ \Omega^{1}_{SO}(BPin^-(2))= & \Z_2^{A^2} \\ \Omega^{2}_{SO}(BPin^-(2))= & 0\\ \Omega^{3}_{SO}(BPin^-(2))= & \Z^{p_1/3} \oplus \Z^{c_1^2} \\ \Omega^{4}_{SO}(BPin^-(2))= & 0.
\end{matrix}\]

For $\tau = \sigma$, we get
\begin{center}
\begin{tikzcd}[row sep=small, column sep = {3em,between origins}]
& 0  \\
& 0 & \Z_2^{A p_1/3} \\
& 0 & 0 & 0 & 0  \\
& 0  & 0 & 0 & 0 & 0 \\
& 0  & 0 & 0 & 0 & 0 & 0\\
q = -1 & 0 & \Z_2^A & \Z^{c_1} & 0 & 0 & \Z_2^{A c_1^2} & 0 \\
& p=0
\end{tikzcd}
\end{center}
(again with no possible differentials.) This gives
\[\begin{matrix}
    \Omega^{-1}_{SO}(BPin^-(2),\sigma) = & 0 \\ \Omega^{0}_{SO}(BPin^-(2),\sigma)= & \Z_2^A \\ \Omega^{1}_{SO}(BPin^-(2),\sigma)= & \Z^{c_1} \\ \Omega^{2}_{SO}(BPin^-(2),\sigma)= & 0 \\ \Omega^{3}_{SO}(BPin^-(2),\sigma)= & 0 \\ \Omega^{4}_{SO}(BPin^-(2),\sigma)= & \Z_2^{Ap_1/3} \oplus \Z_2^{A c_1^2}  .
\end{matrix}\]
The last group is a priori ambiguous because we need to solve the extension problem of the spectral sequence. However, it follows from a theorem of Wall \cite{wall1960} that the oriented bordism spectrum at $p = 2$ is a product of Eilenberg-Maclane spectra, and so all $\Z_2$ extensions, like the one above, split.

We can also fix the ambiguity by using the symmetry breaking long-exact sequence (SBLES) \cite{debray2024longexactsequencesymmetry}. We can use the representation coming from the quotient $Pin^-(2) \to O(2)$ to study the symmetry breaking LES from this symmetry class to the one above. This breaks $Pin^-(2)$ to $\Z_4$. The cobordism groups of $(\bZ_4,\sigma)$ twisted oriented manifolds was computed in low dimensions by \cite{Wan_2020} (they called it $E$-structure). They are
\[\begin{matrix}
    \Omega^0_{SO}(B\bZ_4,\sigma)= & \bZ_2 \\ \Omega^1_{SO}(B\bZ_4,\sigma)= & 0 \\ \Omega^2_{SO}(B\bZ_4,\sigma)= & \bZ_2 \\ \Omega^3_{SO}(B\bZ_4,\sigma)= & 0 \\ \Omega^4_{SO}(B\bZ_4,\sigma)= & \bZ_2^2 \\ \Omega^5_{SO}(B\bZ_4,\sigma)= & \bZ_2.
\end{matrix}\]
We want to do the calculation also for $(Pin^-(2),\sigma)$. We can use the representation coming from the quotient $Pin^-(2) \to O(2)$ to study the symmetry breaking LES from this symmetry class to the one above. We get
\begin{center}
\begin{tikzcd}[row sep=small, column sep = small]
	D & \Omega^{D-2}_{SO}(BPin^-(2)) & \Omega^{D}_{SO}(BPin^-(2),\sigma)& \Omega^{D}_{SO}(B\Z_4,\sigma) \\
	0 & 0 & \Z_2 & \Z_2 \\
	1 & \Z & \Z & 0\\
	2 & 0 & 0 & \Z_2 \\
	3 & \Z_2 & 0 & 0 \\
	4 & 0 & \Omega^{4}_{SO}(BPin^-(2),\sigma) & \Z_2^2 \\
	5 & \Z \oplus \Z &  &  \\
	\arrow[from=2-3, to=2-4]
	\arrow[from=3-2, to=3-3]
	\arrow[from=4-4, to=5-2, out=-20, in=160]
	\arrow[from=6-3, to=6-4]
\end{tikzcd}
\end{center}
Thus we obtain an isomorphism $\Omega^4_{SO}(BPin^-(x),\sigma) = \Omega^4_{SO}(B\Z_4,\sigma)= \Z_2^2$, which resolves the extension problem.

\subsection{$U(2n) \cdot CT$}

Now we study the extension
\[U(2n) \to U(2n) \cdot CT \to \Z_2,\]
defined by
\[U(2n) \cdot CT = (Pin^-(2)\times SU(2n))/\Z_{2n},\]
where $\Z_{2n}$ is the subgroup generated by the product of the $e^{i\pi/n}$ element of $Pin^-(2)$ and the central element $e^{-i\pi/n}I_{2n}$ of $SU(2n)$. The extension is such that $CT$ acts trivially on $SU(2n)$ and squares to the element $-I_{2n}$.

This calculation is very similar to the previous one. The LHSSS has $E_2^{p,q} = H^p(B\bZ_2,H^q(BU(n),\Z)$ where we can choose whether $\Z_2$ acts on the coefficients $\Z$ with twist $\tau$.

For $\tau$ trivial we get
\begin{center}
\begin{tikzcd}[row sep=small, column sep = {3em,between origins}]
& 0 \\
& 0 & 0 \\
& \Z^{c_1^2} \oplus \Z^{c_2} & 0 & \Z_2^{A^2 c_1^2} \oplus \Z_2^{A^2 c_2} \\
& 0 & 0 & 0 & 0 \\
& 0 & \Z_2^{Ac_1} & 0 & \Z_2^{A^3 c_1} & 0 & \Z_2^{A^5 c_1} \\
& 0  & 0 & 0 & 0 & 0 & 0 & 0\\
q = 0 & \Z & 0 & \Z_2^{A^2} & 0 & \Z_2^{A^4} & 0 & \Z_2^{A^6} & 0 & \Z_2^{A^8} \\
& p = 0
	\arrow[from=5-3, to=7-6]
 \arrow[from=5-5, to=7-8]
 \arrow[from=5-7, to=7-10]
\end{tikzcd}
\end{center}
The arrows indicate the differential $d_3(c_1) = A^3$, which comes from the non-trivial extension. This yields
\[\begin{matrix}
    H^0(BU(2n) \cdot CT,\Z)= & \Z \\ H^1(BU(2n) \cdot CT,\Z)= & 0 \\ H^2(BU(2n) \cdot CT,\Z)= & \Z_2^{A^2} \\ H^3(BU(2n) \cdot CT,\Z)= & 0 \\ H^4(BU(2n) \cdot CT,\Z)= & \Z^{c_1^2} \oplus \Z^{c_2} \\ H^5(BU(2n) \cdot CT,\Z)= & 0 \\ H^6(BU(2n) \cdot CT,\Z)= & \Z_2^{A^2 c_1^2} \oplus \Z_2^{A^2 c_2}
\end{matrix}\]
For $\tau = \sigma$ we get
\begin{center}
\begin{tikzcd}[row sep=small, column sep = {3em,between origins}]
& 0 \\
& 0 & 0 \\
& 0 & \Z_2^{A c_1^2} \oplus \Z_2^{Ac_2} & 0 \\
& 0 & 0 & 0 & 0 \\
& \Z^{c_1} & 0 & \Z_2^{A^2 c_1} & 0 & \Z_2^{A^4 c_1} & 0\\
& 0  & 0 & 0 & 0 & 0 & 0 & 0\\
q = 0 & 0 & \Z_2^A & 0 & \Z_2^{A^3} & 0 & \Z_2^{A^5} & 0 & \Z_2^{A^7} \\
& p = 0
	\arrow[from=5-2, to=7-5]
 \arrow[from=5-4, to=7-7]
 \arrow[from=5-6, to=7-9]
\end{tikzcd}
\end{center}
This yields
\[\begin{matrix}
    H^0(BU(2n) \cdot CT,\Z^\sigma)= & 0 \\
    H^1(BU(2n) \cdot CT,\Z^\sigma)= & \Z_2^A \\ 
    H^2(BU(2n) \cdot CT,\Z^\sigma)= & \Z^{2 c_1} \\
    H^3(BU(2n) \cdot CT,\Z^\sigma)= & 0 \\
    H^4(BU(2n) \cdot CT,\Z^\sigma)= & 0 \\
    H^5(BU(2n) \cdot CT,\Z^\sigma)= & \Z_2^{A c_1^2} \oplus \Z_2^{A c_2} \\
    H^6(BU(2n) \cdot CT,\Z^\sigma)= &0.
\end{matrix}\]
The $\tau$-twisted oriented cobordism groups of $U(2n) \cdot CT$ we need can now be computed by the Atiyah-Hirzebruch spectral sequence (AHSS), which has $E_2^{p,q} = H^p(BU(2n) \cdot CT,\Omega^q_{SO} \otimes \Z^\tau)$. For $\tau$ trivial we get
\begin{center}
\begin{tikzcd}[row sep=small, column sep = {3em,between origins}]
& 0  \\
& \Z^{p_1/3} & 0 \\
& 0 & 0 & 0  \\
& 0  & 0 & 0 & 0 \\
& 0  & 0 & 0 & 0 & 0 \\
q = -1 & \Z & 0 & \Z_2^{A^2} & 0 & \Z^{c_1^2} \oplus \Z^{c_2} & 0 \\
& p=0
\end{tikzcd}
\end{center}
Which gives
\[\begin{matrix}
    \Omega^{-1}_{SO}(BU(2n) \cdot CT) = & \Z \\
    \Omega^{0}_{SO}(BU(2n) \cdot CT)= & 0 \\
    \Omega^{1}_{SO}(BU(2n) \cdot CT)= & \Z_2^{A^2} \\
    \Omega^{2}_{SO}(BU(2n) \cdot CT)= & 0\\
    \Omega^{3}_{SO}(BU(2n) \cdot CT)= & \Z^{p_1/3} \oplus \Z^{c_1^2} \oplus \Z^{c_2} \\
    \Omega^{4}_{SO}(BU(2n) \cdot CT)= & 0.
\end{matrix}\]
For $\tau = \sigma$, we get
\begin{center}
\begin{tikzcd}[row sep=small, column sep = {4em,between origins}]
& 0  \\
& 0 & \Z_2^{A p_1/3} \\
& 0 & 0 & 0 & 0  \\
& 0  & 0 & 0 & 0 & 0 \\
& 0  & 0 & 0 & 0 & 0 & 0\\
q = - 1 & 0 & \Z_2^A & \Z^{c_1} & 0 & 0 & \Z_2^{A c_1^2} \oplus \Z_2^{A c_2} & 0 \\
& p = 0
\end{tikzcd}
\end{center}
This gives
\[\begin{matrix}
    \Omega^{-1}_{SO}(BU(2n) \cdot CT,\sigma) = & 0 \\
    \Omega^{0}_{SO}(BU(2n) \cdot CT,\sigma)= & \Z_2^A \\
    \Omega^{1}_{SO}(BU(2n) \cdot CT,\sigma)= & \Z^{c_1} \\
    \Omega^{2}_{SO}(BU(2n) \cdot CT,\sigma)= & 0 \\
    \Omega^{3}_{SO}(BU(2n) \cdot CT,\sigma)= & 0 \\
    \Omega^{4}_{SO}(BU(2n) \cdot CT,\sigma)= & \Z_2^{Ap_1/3} \oplus \Z_2^{A c_1^2} \oplus \Z_2^{A c_2} .
\end{matrix}\]
Again the last group is a product because of the splitting of oriented bordism at $p = 2$.

We could also resolve this extension problem by considering a different subgroup $Pin^-(2) \to U(2n) \cdot CT$ where the $U(1)$ subgroup of $Pin^-(2)$ maps to a $U(1)$ subgroup of $SU(2n)$ (rather than the center of $U(2n)$). $c_2 \in H^4(BSU(2n),\Z)$ pulls back to $c_1^2 \in H^4(BU(1),\Z)$ under this map. The AHSS is functorial under this pullback, so since the extension is trivial for $Pin^-(2)$, it is also trivial for $U(2n) \cdot CT$.

\subsection{$Pin^-(2) \rtimes \Z_2^C$}

Now we want to add another $\Z_2$ symmetry, charge conjugation $C$ into the mix. This acts as complex conjugation on the $U(2n)$ matrices but commutes with $CT$, with no extension. So the groups we will study have the split form $G \rtimes \Z_2^C$, where $G = Pin^-(2)$ or $G = U(2n) \cdot CT$. We study the AHSS for
\[G \to G \rtimes \Z_2^C \to \Z_2^C\]
with
\[E_2^{p,q} = H^p(B\Z_2,\Omega^q_{SO}(BG,\tau))\]
for arbitrary twist $\tau$.

For $\tau$ trivial, $G = Pin^-(2)$ we get
\begin{center}
\begin{tikzcd}[row sep=small, column sep = {3em,between origins}]
& 0 \\
& \Z^{p_1/3} \oplus \Z^{c_1^2} &0 \\
& 0 & 0  & 0 \\
& \Z_2^{A^2}  & \Z_2^{A^2 C} & \Z_2^{A^2 C^2} & \Z_2^{A^2 C^3}\\
& 0  & 0 & 0 & 0 & 0 \\
q = - 1& \Z & 0 & \Z_2^{C^2} & 0 & \Z_2^{C^4} & 0 & \Z_2^{C^6}\\
& p = 0
\end{tikzcd}
\end{center}
There are possible differentials here which we don't know how to rule out.

For $\tau = \sigma$, $G = Pin^-(2)$ we get
\begin{center}
\begin{tikzcd}[row sep=small, column sep = {4em,between origins}]
& \Z_2^{Ap_1/3} \oplus \Z_2^{Ac_1^2} \\
& 0 & 0  & 0 \\
& 0  & 0 & 0 & 0\\
& 0  & \Z_2^{C c_1} & 0 & \Z_2^{C^3 c_1} & 0 \\
q = 0 & \Z_2^A & \Z_2^{AC} & \Z_2^{AC^2} & \Z_2^{AC^3} & \Z_2^{AC^4} & \Z_2^{AC^5} \\
& p = 0
\end{tikzcd}
\end{center}
There are possible differentials landing in $\Z_2^{AC^{2n+1}}$ but we can show these must be zero. In particular, $\mathbb{RP}^{2n+1}$ is an oriented manifold with a $\bZ_2$ bundle $C$ with $\int_{\mathbb{RP}^{2n+1}} C^{2n+1} = 1$ mod 2. This can be extended to any $G \rtimes \Z_2^C$ connection by taking the $G$ part to be trivial. Thus, we obtain
\[\begin{matrix}
    \Omega^{0}_{SO}(B[Pin^-(2) \rtimes \Z_2^C],\sigma) = & \Z_2^A \\ \Omega^{1}_{SO}(B[Pin^-(2) \rtimes \Z_2^C],\sigma)= & \Z_2^{AC} \\
    \Omega^{2}_{SO}(B[Pin^-(2) \rtimes \Z_2^C],\sigma)= & \Z_2^{A C^2} \oplus \Z_2^{C c_1} \\
    \Omega^{3}_{SO}(B[Pin^-(2) \rtimes \Z_2^C],\sigma)= & \Z_2^{AC^3}\\
    \Omega^{4}_{SO}(B[Pin^-(2) \rtimes \Z_2^C],\sigma)= & \Z^{Ap_1/3} \oplus \Z^{Ac_1^2} \oplus \Z_2^{C^3 c_1} \oplus \Z_2^{AC^4}
\end{matrix}\]

\subsection{$U(2n) \cdot (CT \times C)$}\label{appendixU2result}

Now we want to do the calculation for a group $G = (U(2n) \cdot CT) \rtimes \bZ_2^C$, where $C$ commutes with $CT$ while acting as complex conjugation on the $U(2n)$ matrices. We can mix the LHS and AHSS spectral sequences to obtain one with $E_2^{p,q} = H^p(B\Z_2,\Omega^q_{SO}(BU(2n) \cdot CT,\tau))$ for arbitrary twist $\tau$. For $\tau = \sigma$, we get
\begin{center}
\begin{tikzcd}[row sep=small, column sep = {3em,between origins}]
& \Z_2^{Ap_1/3} \oplus \Z_2^{Ac_1^2} \oplus \Z_2^{Ac_2} \\
& 0 & 0  & 0 \\
& 0  & 0 & 0 & 0\\
& 0  & \Z_2^{C c_1} & 0 & \Z_2^{C^3 c_1} & 0 \\
q = 0 & \Z_2^A & \Z_2^{AC} & \Z_2^{AC^2} & \Z_2^{AC^3} & \Z_2^{AC^4} & \Z_2^{AC^5} \\
& p = 0
\end{tikzcd}
\end{center}
Again there are possible differentials landing in $\Z_2^{C^{2n+1}}$ but by the argument in the ${\rm Pin}^-(2) \rtimes \Z_2^C$ case above these are trivial.

Thus, we obtain
\begin{equation}
    \label{eqnanomclassification}\begin{matrix}
    \Omega^{0}_{SO}(B[(U(2n) \cdot CT) \rtimes \Z_2^C],\sigma) = & \Z_2^A \\ \Omega^{1}_{SO}(B[(U(2n) \cdot CT) \rtimes \Z_2^C],\sigma)= & \Z_2^{AC} \\
    \Omega^{2}_{SO}(B[(U(2n) \cdot CT) \rtimes \Z_2^C],\sigma)= & \Z_2^{A C^2} \oplus \Z_2^{C c_1} \\
    \Omega^{3}_{SO}(B[(U(2n) \cdot CT) \rtimes \Z_2^C],\sigma)= & \Z_2^{AC^3}\\
    \Omega^{4}_{SO}(B[(U(2n) \cdot CT) \rtimes \Z_2^C],\sigma)= & \Z^{Ap_1/3} \oplus \Z^{Ac_2} \oplus \Z^{Ac_1^2} \oplus \Z_2^{C^3 c_1} \oplus \Z_2^{AC^4}.
    \end{matrix}
\end{equation}
Our case of interest for $N_f = 2$ QED$_3$ is $\Omega^4_{SO}(\cdots)$ for $n = 1$.

It is useful to know for calculations also the restriction of these anomalies to the group $O(2)$ generated by the matrices
\[\begin{pmatrix}
    e^{2\pi i \theta} & 0 \\ 0 & I_{2n-1}
\end{pmatrix}\]
and $C$. The cohomology $H^k(BO(2),\Z)$ is computed in \cite{browncohomologyOn}, where it is shown (see theorem 1.6) to be generated by the 2-torsion classes $w_1^2$ and $w_1 w_2$, as well as the non-torsion class $p_1$, subject to some complicated relations. The upshot is that the low degree groups are
\begin{equation}
    \begin{matrix}
    H^1(BO(2),\Z) = & 0 \\
    H^2(BO(2),\Z) = & \Z_2^{w_1^2} \\
    H^3(BO(2),\Z) = & \Z_2^{w_1 w_2} \\
    H^4(BO(2),\Z) = & \Z^{p_1} \oplus \Z_2^{w_1^4} \\
    H^5(BO(2),\Z) = & \Z_2^{w_1^3 w_2}.
    \end{matrix}
\end{equation}
We find $c_1(U(2))$ restricts to $w_2$ and $C$ may be identified with $w_1$. This restriction therefore maps $Ap_1/3$, $Ac_2$, $Ac_1^2$, and $AC^4$ to zero in $\Omega^4_{SO}(BO(2))$, and maps $C^3 c_1$ to $w_1^3 w_2$. This class is non-trivial, and an example test manifold is $\mathbb{RP}^3 \times S^1$ with $C = x$, where $x$ is the generator of $H^1(\mathbb{RP}^3,\Z_2)$, and twisted Euler class $e = x w \in H^2(\mathbb{RP}^3 \times S^1, \Z^x)$, where $w$ is the generator of $H^1(S^1,\Z) = \Z$.

\section{Equivariant Cohomology Calculations}

\subsection{$S^2//U(2)$}\label{appendixU2equivS2coh}

Suppose we have a theory of an $S^2$ sigma field $n$ and a global $U(2)$ symmetry which acts on $S^2$ by the real triplet representation $U(2) \to SO(3)$. Turning on a background field for this $U(2)$ symmetry means that $n$ becomes a section of the $S^2$ bundle associated to the $U(2)$ gauge bundle. The data of these two bundles over a space $M$ is equivalent to a homotopy class of maps $M \to S^2//U(2)$, the latter space being the homotopy quotient of $S^2$ by $U(2)$. This is the $S^2$ bundle over $BU(2)$ associated to the tautological $U(2)$ bundle by the $U(2)$ action on $S^2$:
\[
\begin{tikzcd}
    S^2 \arrow[r, "i"] & S^2//U(2) \arrow[d, "\pi"] \\
    & BU(2)
\end{tikzcd}.
\]
Expressions like \eqref{covOmega} can be understood as representation cohomology classes on $S^2//U(2)$. That expression in particular pulls back by $i^*$ to the volume form on the fiber $S^2$.

Since $S^2$ has a transitive action by $U(2)$, it is homotopy equivalent to $BU(1)^2$, where $U(1)^2$ is the stabilizer of any chosen point on $S^2$. One can think of this as the ability to gauge fix the data of the background $U(2)$ gauge field and section of the associated $S^2$ bundle by choosing $U(2)$ functions making this $S^2$ section constant.

The integer cohomology of $S^2//U(2)$ is thus simple to compute, with two free generators $x,y$ in degree 2, which we can identify with the Chern classes of the two unbroken $U(1)$'s. These are simple to relate to cohomology classes on $U(2)$ by the Whitney formula. In particular, we have
\[\pi^* c_1(U(2)) = x + y~, \qquad
\pi^* c_2(U(2)) = x y~.\]
To relate these also to the cohomology of $S^2$ we can use the Serre spectral sequence, with $E_2^{p,q} = H^p(BU(2),H^q(S^2))$:
\begin{center}
\begin{tikzcd}[row sep=small, column sep = {3em,between origins}]
&0  \\
& 0 & 0\\
& 0 & 0  & 0 \\
& \Z^{\Omega_2} & 0 & \Z^{\Omega_2 c_1} & 0\\
& 0  & 0 & 0 & 0 & 0 \\
q = 0 & \Z & 0 & \Z^{c_1} & 0 & \Z^{c_1^2} \oplus \Z^{c_2} & 0 \\
& p = 0
\end{tikzcd}
\end{center}
There are no possible differentials in this spectral sequence, which means that expressions such as $\Omega_2$ extend to cocycles $\tilde \Omega_2^{U(2)}$ on the whole space $S^2//U(2)$. This is the fully $U(2)$-equivariantization which in terms of the $SO(3)$-equivariantization in \eqref{covOmega} is $\tilde \Omega_2 - \frac{dA_m}{2\pi}$.

We want to identify these cohomology classes with combinations of the generators of $S^2//U(2) = BU(1)^2$, $x$ and $y$. We have already identified
\[\pi^*c_1 = x + y,\]
by the Whitney sum formula. We also have
\[\pi^*c_2 = xy\]
by the same. $\tilde \Omega_2^{U(2)}$ meanwhile has to yield another integer generator of $\Z^x \oplus \Z^y = H^2(BU(1)^2,\Z)$, and so
\[\tilde \Omega_2^{U(2)} = x + k(x+y)\]
for some integer $k$.

The integrand in the second expression in \eqref{c2identity} can thus be identified with 
\[(w_2(T\cM) + x+y) \cup (x + k(x+y)) = x \cup y = \pi^* c_2(U(2)) \mod 2~,\]
which finishes the derivation of the anomaly given there.

\subsection{$S^3//U(2)$}\label{appendixS3withU2action}

Suppose we consider now $S^3$ with $U(2)$ acting by the fundamental representation on the unit sphere in $\bC^2$. The homotopy quotient in this case is $S^3//U(2) = BU(1)$, so the Serre spectral sequence with $E_2^{p,q} = H^p(BU(2),H^q(S^3))$ has a differential:
\begin{center}
\begin{tikzcd}[row sep=small, column sep = {3em,between origins}]
& 0 \\
& \Z^{\Omega_3} & 0 \\
& 0 & 0 & 0 \\
& 0  & 0 & 0 & 0  \\
q = 0 & \Z & 0 & \Z^{c_1} & 0 & \Z^{c_1^2} \oplus \Z^{c_2}  \\
& p = 0
\arrow[from=2-2, to=5-6]
\end{tikzcd}
\end{center}
This differential must be nonzero to reproduce $H^3(S^3//U(1),\bZ) = H^3(BU(1),\bZ) = 0$. It means that the equivariantization $\tilde \Omega_3$ of $\Omega_3$ is not closed, but instead satisfies
\[d\tilde \Omega_3 = \pi^* c_2(U(2)),\]
which is well-known to be the Euler class of the $S^3$ bundle over $U(2)$, and equal to the top Chern class, $c_2(U(2))$ \cite{bott2013differential}.

\subsection{$S^2//SO(3)$}\label{SO3equivappendix}

Another interesting case is $SO(3)$ acting on $S^2$ via the vector representation. The homotopy quotient is again $S^2//SO(3) = BU(1)$. The Serre spectral sequence is $E_2^{p,q}=H^p(BSO(3),H^q(S^2,\Z))$ which reads
\begin{center}
\begin{tikzcd}[row sep=small, column sep = {3em,between origins}]
& 0 \\
& 0 & 0 \\
& \Z^{\Omega_2} & 0 & 0 \\
& 0  & 0 & 0 & 0  \\
q = 0 & \Z & 0 & 0 & \Z_2^{w_3} & \Z^{p_1} & 0  \\
& p = 0
\arrow[from=3-2, to=5-5]
\end{tikzcd}
\end{center}
The differential has to be there to yield $H^3(S^2//SO(3),\bZ) = H^3(BU(1),\bZ) = 0$. This shows that the equivariantized $\Omega_2$ in \eqref{covOmega} can have half-integral periods, equal to $\frac{1}{2} w_2(SO(3))$ mod 1, since $\frac{d w_2}{2} = w_3$.

\section{Invertibility of Level-1 Spin$^c$ Chern-Simons theory}\label{spincchernsimonsappendix}

Let $A$ be a Spin$^c$ connection on a 3-manifold $\cM_3$. The bordism group $\Omega_3^{{\rm Spin}^c} = 0$, so we can attempt to define Chern-Simons invariants by extending $A$ to a 4-manifold $\cM_4$ with $\partial \cM_4 = \cM_3$. Normalizing relative to the ordinary $U(1)$ case, a level $k$ term would be defined as
\[\exp\left( \frac{ik}{4\pi} \int_{\cM_4} dA \wedge dA \right) \,.\]
As written, this will depend on the choice of $\cM_4$, since $\oint_2 dA = \pi \oint_2 w_2(T\cM) + 2\pi \bZ$. However, the Atiyah-Singer index theorem for closed 4-manifolds says that
\[\oint_{\cM_4} \left(\hat A(R) + \frac{1}{2} \frac{dA}{2\pi} \wedge \frac{dA}{2\pi} \right) \in \bZ\,,\]
since this integral is the index of the $A$-twisted Dirac operator on $\cM_4$. Here $\hat A(R)$ is the A-roof genus, which depends on the metric curvature $R$ on $\cM_4$. It satisfies
\[\oint_{\cM_4} \hat A(R) = \frac{\sigma(\cM_4)}{8}\,,\]
where $\sigma(\cM_4)$ is the signature of $\cM_4$. So a proper definition of the level $k$ Spin$^c$ term is
\[\exp\left( \frac{ik}{4\pi} \int_{\cM_4}  dA \wedge dA + 8\pi^2 \hat A(R) \right)\,.\]
When $k$ is divisible by 4, the $\hat A(R)$ term becomes a separate gravitational Chern-Simons term at level $k/4$ which can be split off. See also appendix A of \cite{Seiberg:2016rsg}.

We want to study a theory of a dynamical Spin$^c$ structure with this term with $k = 1$. One way to regularize this is to fix a background Spin$^c$ structure $A_0$ and write $A = A_0 + a$, where $a$ is a dynamical, ordinary $U(1)$ gauge field. The path integral weight becomes
\[\exp\left( \frac{i}{4\pi} \int_{\cM_4} da \wedge da + 2 da \wedge dA_0 + dA_0 \wedge dA_0 + 8\pi^2 \hat A(R) \right)\,.\]
The path integral over $a$ can now be treated in the usual way and will not depend on the choice of $A_0$ (it is thus a bosonic theory).

Consider a 3-torus $\cM_3 = T^3$ with flat metric and choose $A_0$ to be the all anti-periodic spin structure on $T^3$. This extends along with $a$ to a solid torus $D^2 \times T^2$ with flat metric. The path integral weight thus becomes simply
\[\exp\left( \frac{i}{4\pi} \int_{\cM_4} da \wedge da \right)\,,\]
which is the same as ordinary spin $U(1)_1$ Chern-Simons theory with this spin structure. The TQFT partition function on $T^3$ is therefore 1, because $U(1)_1$ is an invertible spin TQFT \cite{Witten:2003ya,Seiberg:2016gmd,Hsin:2016blu}. Thus the Spin$^c$ $U(1)_1$ has a unique ground state on $T^2$. By the main theorem of \cite{Schommer_Pries_2018}, it follows the whole theory is invertible.

\bibliographystyle{JHEP}
\bibliography{QED3}

\end{document}